\pdfoutput=1
\documentclass[runningheads]{llncs}

\usepackage[frozencache,cachedir=minted-cache]{minted}

\usepackage{booktabs}   
\usepackage{subcaption} 

\usepackage{soul}

\usepackage{array}
\newcolumntype{H}{>{\setbox0=\hbox\bgroup}c<{\egroup}@{}}

\usepackage{xr}

\usepackage{float}
\usepackage{multicol}
\usepackage{fixltx2e}
\usepackage{todonotes}
\usepackage{amsmath}
\usepackage{xspace}
\usepackage{refs}
\usepackage{galois}
\usepackage{xcolor}
\usepackage{multirow}
\usepackage[framemethod=tikz]{mdframed}

\usepackage{wrapfig}
\usepackage[export]{adjustbox}
\usepackage{stfloats}
\usepackage{hyperref}
\hypersetup{
	colorlinks,
	linkcolor={red!50!black},
	citecolor={blue!50!black},
	urlcolor={blue!80!black}
}

\usepackage{cleveref}
\crefname{table}{Tab.}{Tab.}
\Crefname{table}{Table}{Tables}
\crefname{section}{Sect.}{Sect.}
\Crefname{section}{Section}{Sections}
\crefname{listing}{\lstlistingname}{\lstlistingname}
\Crefname{listing}{Listing}{Listings}
\crefname{algocf}{Alg.}{Algs.}
\Crefname{algocf}{Algorithm}{Algorithms}
\creflabelformat{equation}{#2#1#3}

\newcommand{\labelline}[1]{\label[line]{#1}\hypertarget{#1}{}}
\newcommand{\refline}[1]{\hyperlink{#1}{Line~\ref*{#1}}}

\usepackage{tikz}
\usetikzlibrary{shapes}
\usetikzlibrary{patterns}
\usetikzlibrary{circuits}
\usetikzlibrary{fit}
\usetikzlibrary{matrix}
\usetikzlibrary{arrows}
\usetikzlibrary{calc}
\usetikzlibrary{positioning}
\tikzset{>=stealth',every on chain/.append style={join}, every join/.style={->}}

\definecolor{pale}{rgb}{.97, .95, .7}
\newcommand*\circled[1]{\tikz[baseline=(char.base)]{
            \node[shape=circle,draw,inner sep=2pt,fill=pale] (char) {#1};}}

\usepackage{makecell}
\usepackage{stmaryrd}

\usepackage{etoolbox}

\newrobustcmd*{\nsquare}[1]{\tikz{\filldraw[draw=#1,fill=#1] (0,0)
    rectangle (0.2cm,0.2cm);}}

\newrobustcmd*{\ncircle}[1]{\tikz{\filldraw[draw=#1,fill=#1] (0,0) circle [radius=0.1cm];}}

\newrobustcmd*{\ntriangle}[1]{\tikz{\filldraw[draw=#1,fill=#1] (0,0) --
    (0.2cm,0) -- (0.1cm,0.2cm);}}

\usepackage{listings}
\usepackage{listingsutf8}

\definecolor{darklava}{rgb}{0.58, 0.1, 0.9}

\usepackage{ amsfonts }

\usepackage{comment}
\usepackage[algo2e,linesnumbered,ruled,vlined,resetcount]{algorithm2e}
\usepackage{textcomp}

\SetKwComment{Comment}{$\triangleright$\ }{}

\SetCommentSty{xCommentSty}

\usepackage{color, colortbl}
\usepackage{xifthen}
\usepackage{pifont}
\usepackage{diagbox}
\usepackage[normalem]{ulem}

\newcommand{\tool}{\textsc{Amurth2}\xspace}
\newcommand{\amurth}{\textsc{Amurth}\xspace}
\newcommand{\func}[1]{\textsc{#1}}
\newcommand{\abst}[2]{\ensuremath{#1^\sharp_{#2}}}
\newcommand{\pex}{\ensuremath{E^{+}}}
\newcommand{\nex}{\ensuremath{E^{-}}}

\newcommand{\fsyn}[1][]{\ensuremath{$f_{E}^\sharp$}}
\newcommand{\fSharpSyn}[1][]{
  \ifthenelse{\isempty{#1}}
             {f_{E}^\sharp}
             {f_{E}^\sharp\left{#1}\right}
}

\newcommand{\AbsDomain}{A}

\newcommand{\Lang}{{\mathcal{L}}}

\newcommand{\safe}{\textsc{SAFE}\xspace}
\newcommand{\jsai}{\textsc{JSAI}\xspace}
\newcommand{\no}{\ensuremath{\mathcal{NO}}\xspace}
\newcommand{\nos}{\ensuremath{\mathcal{NOS}}\xspace}

\newcommand{\safestr}{SAFE\textsubscript{str}\xspace}
\newcommand{\csdom}{\ensuremath{\mathcal{CS}}\xspace}
\newcommand{\ssk}{\ensuremath{\mathcal{SS}_k}\xspace}

\usepackage{enumitem}
\setlist{leftmargin=*}
\setlist{nosep,leftmargin=\parindent}
\setlist[description]{leftmargin=\parindent,labelindent=\parindent}

\definecolor{crimson}{HTML}{C90016}
\definecolor{tmagenta}{HTML}{FF00FF}
\definecolor{tred}{HTML}{FF0000}
\definecolor{posex}{HTML}{0ca765}
\definecolor{bottlegreen}{rgb}{0.0,0.42,0.31}

\newcommand{\D}{\ensuremath{\mathcal{D}}}
\newcommand{\Omit}[1]{}

\newcommand{\concat}{\texttt{concat}\xspace}
\newcommand{\tolower}{\texttt{toLower}\xspace}
\newcommand{\toupper}{\texttt{toUpper}\xspace}
\newcommand{\trim}{\texttt{trim}\xspace}
\newcommand{\contains}{\texttt{contains}\xspace}
\newcommand{\charat}{\texttt{charAt}\xspace}

\newcommand{\eqdef}{\operatorname{\,{=_{\textit{df}}}\,}}

\newcommand{\args}{a_1, a_2, \dots, a_n}
\newcommand{\dfuns}{\absr{f}_1, \absr{f}_2, \dots, \absr{f}_n}

\newcommand{\twr}[1]{{{\color{purple}{T: #1}}}}

\newcommand{\pankaj}[1]{{\color{blue}{#1}}}

\newcommand{\dipt}[3]{\ensuremath{\mathtt{#1^{\sharp \textsf{D}}(\langle #2\rangle, \langle #3\rangle)}}}
\newcommand{\repo}[2]{\ensuremath{\mathtt{#1^{\sharp \textsf{R}}(\langle #2\rangle)}}}
\newcommand{\rept}[3]{\ensuremath{\mathtt{#1^{\sharp \textsf{R}}(\langle #2\rangle, \langle #3\rangle)}}}

\newcommand{\abs}[1]{{#1}^\sharp}
\newcommand{\absd}[1]{{#1}^{\sharp \textsf{D}}}
\newcommand{\absr}[1]{{#1}^{\sharp \textsf{R}}}
\renewcommand{\b}[1]{\langle #1 \rangle}
\newcommand{\abstr}[2]{\ensuremath{#1^{\sharp \textsf{R}}_{#2}}}

\begin{document}

\title{Synthesizing Abstract Transformers for Reduced-Product Domains}


\author{Pankaj Kumar Kalita\inst{1}\orcidID{0000-0001-5826-0030} \and Thomas Reps\inst{2}\orcidID{0000-0002-5676-9949} \and Subhajit Roy\inst{1}\orcidID{0000-0002-3394-023X}}
\institute{Indian Institute of Technology Kanpur \\\email{\{pkalita,subhajit\}@cse.iitk.ac.in} \and University\ of Wisconsin-Madison\\ \email{reps@cs.wisc.edu}}
\authorrunning{P.K.\ Kalita et al.}





\maketitle              

\begin{abstract}
Recently, we showed how to apply program-synthesis techniques to create abstract transformers in a user-provided domain-specific language (DSL) $\Lang$ (i.e., ``$\Lang$-transformers'').
However, we found that the algorithm of Kalita et al.\ does not succeed when applied to reduced-product domains: the need to synthesize transformers for all of the domains simultaneously blows up the search space.

\hspace{1.5ex}
Because reduced-product domains are an important device for improving the precision of abstract interpretation, in this paper, we propose an algorithm to synthesize reduced $\Lang$-transformers $\b{\dfuns}$ for a product domain $A_1 \times A_2 \times \dots \times A_n$, using multiple DSLs: $\Lang$ $= \langle \Lang_1, \Lang_2, \ldots, \Lang_n \rangle$. Synthesis of reduced-product transformers is quite challenging: first, the synthesis task has to tackle an increased ``feature set'' because each component transformer now has access to the abstract inputs from all component domains in the product.
Second, to ensure that the product transformer is maximally precise, the synthesis task needs to arrange for the component transformers to cooperate with each other.

\hspace{1.5ex}
We implemented our algorithm in a tool, \tool, and used it to synthesize abstract transformers for two product domains---SAFE and JSAI---available within
the \safestr framework for JavaScript program analysis.
For four of the six operations supported by \safestr,
\tool synthesizes more precise abstract transformers than the manually written ones available in \safestr.
\end{abstract}

\section{Introduction}
\label{sec:intro}

Abstract interpretation~\cite{POPL:CC77} is a program-verification methodology that interprets programs on \textit{abstract states} to reason about program correctness. An abstract state represents a potentially unbounded number of concrete states, thereby enabling reasoning about a set of states \emph{en masse}. The abstract states are defined in carefully constructed \textit{abstract domains}; an \textit{abstraction} function ($\alpha$) and a \textit{concretization} function ($\gamma$) map a set of concrete values to an \textit{abstract} value and back (respectively). For the reasoning to be sound, the $\alpha$ and $\gamma$ functions must form a Galois connection~\cite{POPL:CC77}. 

One of the primary challenges to building an abstract interpretation framework is defining \textit{abstract transformers} that provide abstract semantics to every concrete operation available in the source language. The abstract transformers ``lift" the computation from the concrete domain to an abstract domain, enabling reasoning over a potentially unbounded number of states. However, designing sound and precise abstract transformers is challenging because even simple concrete operations can have quite non-trivial abstract transformers. Abstract-interpretation engines have exhibited bugs in their abstract transformers~\cite{TACAS:SAFEstr17}, 
which raises questions about the trustworthiness of verification endeavours.

Kalita et al.\ \cite{amurth} introduced the problem of \emph{transformer synthesis modulo a domain-specific language (DSL)} for a single abstract domain.
Given operation $\textit{op}$ and abstract domain $\AbsDomain$, their method creates an abstract transformer for $\textit{op}$ over $\AbsDomain$, \emph{expressed in DSL $\Lang$}---what they call an ``$\Lang$-transformer (for $\textit{op}$ over $\AbsDomain$).'' Their algorithm is guaranteed to return \emph{a} \emph{best} $\Lang$-transformer. That is, among all $\Lang$-transformers for $\textit{op}$ over $\AbsDomain$, there is no other $\Lang$-transformer that is strictly more precise than the one obtained by their algorithm. However, there may be other $\Lang$-transformers that are incomparable to the one obtained by the algorithm, which is why one says that the algorithm creates ``\emph{a} best $\Lang$-transformer.''

Instead of single domains, running abstract interpretation on a combination of multiple \textit{component domains} $A_i$, that is, interpreting a program within a \textit{product domain} $A_1 \times A_2 \times \dots \times A_n$, is one of the primary approaches to improving the precision of a static-analysis tool.
The abstract values in a product domain $A_1 \times A_2 \times \dots \times A_n$ are tuples $\b{\args}$ over the component domains, such that $a_i\in A_i$.
The answer obtained using a product domain is at least as precise as the answer obtained from any of the individual component domains $A_i$ (and may be more precise), because a concrete value $c$ is excluded from the product domain's answer $\langle a_1, a_2, \ldots, a_n \rangle$ if, in any component domain $A_i$, $c \notin \gamma(a_i)$.

To enable interpretation on product domains, one can design \textit{reduced transformers}, $\b{\dfuns}$,
where---to obtain more precise answers---each component-domain transformer $\absr{f}_i$ is provided access to the abstract values from all the other domains.
While designing transformers for single domains is challenging, designing reduced transformers for product domains is more so. First, each component-domain transformer $\absr{f}$ is provided access to the abstract values from all component domains, thereby increasing the \textit{feature space} for the synthesis task.
Second, the component-domain transformers cannot be synthesized independently---all the component-domain transformers must \textit{cooperate} with each other to produce the maximally precise reduced abstract value. One approach to synthesizing such reduced products is to apply \amurth~\cite{amurth} directly on the product domain. However, this approach is not practical because it requires all the component domain transformers to be synthesized in one second-order query.
The need to synthesize transformers for all of the domains simultaneously blows up the search space tremendously.

In this paper, we provide a practical algorithm to automatically synthesize best $\Lang$-transformers $\b{\dfuns}$ for product domains.
The transformers are expressed in a user-provided domain-specific language $\Lang = \b{\Lang_1, \Lang_2, \dots, \Lang_n}$, where $\Lang_i$ is the domain-specific language to express the component transformer $\absr{f_i}$ corresponding to  domain $A_i$.
We implemented our algorithm in a tool, \tool, that is capable of synthesizing non-trivial reduced-product transformers within reasonable time.
Because \tool synthesizes the component-domain transformers one at a time, it scales much better than directly applying \amurth on the product lattice. For example, \tool could synthesize the \texttt{add} operation of a product domain over odd-intervals and even-intervals (explained in \Cref{sec:back:prod_dom}) in about half an hour, whereas \amurth did not succeed in 10 hours.

We demonstrated the power of \tool by using it to create some reduced abstract transformers for the 
(admittedly artificial) product domain of even-intervals and odd-intervals that we use as a running example in the paper.
As a more important test, we then applied \tool
to synthesize real-world transformers for the \safestr verification framework~\cite{TACAS:SAFEstr17}, designed to detect vulnerabilities in JavaScript programs.
We used \tool to design reduced transformers for two product domains used in \safestr for analyzing string-valued data---SAFE~\cite{FOOL:LWJCR12} and JSAI~\cite{DBLP:conf/sigsoft/KashyapDKWGSWH14}.
Perhaps due to the difficulty of designing reduced transformers,
\safestr only used direct-product transformers---a simple aggregation of the transformers for the component domains---for five of the six string operations supported.
Overall, \tool synthesized more precise transformers than what is used in \safestr for four of the six concrete operations supported by the framework, for both the SAFE and JSAI product domains.

The primary contributions of this work include:
\begin{itemize}
  \item
    We propose a practical algorithm for synthesizing best $\Lang$-transformers for reduced-product domains.
  \item
    We implemented our algorithm in a tool, \tool, that synthesizes such transformers in a reasonable amount of time.
  \item
    We demonstrated the capabilities of \tool by synthesizing best reduced $\Lang$-transformers for transformers available in the \safestr verification framework.
    The transformers synthesized by \tool were found to be more precise than the transformers available in \safestr in many cases.
\end{itemize}

\vspace{10pt}
 The rest of the paper is structured as follows: \Cref{sec:background} gives background on abstract interpretation and product domains.
 \Cref{sec:overview} formulates the problem that we address; articulates the challenges that we face; presents an overview of our approach; and illustrates how the algorithm works, using an example.
 \Cref{sec:algorithm} presents the main algorithm to synthesize abstract transformers for reduced-product domains. \Cref{sec:case_studies} contains the case-studies and our experimental results showing efficacy of \tool. 

The \tool artifact is available in Zenodo~\cite{amurth2artifact}.

\section{Background}
\label{sec:background}

\subsection{Abstract Domains and Transformers} 

An abstract domain is a value-space in which each element describes a (potentially infinite) set of concrete values.
For example, the \textit{interval abstract domain} consists of abstract values of the form $[l,r]$, which represents the set of concrete values $\{x \mid l \leq x \leq r\}$.
Applying abstraction function $\alpha(\{4,6,9\})$ produces the abstract value $[4, 9]$.
However, the concretization function
$\gamma$ applied to the interval $[4,9]$, results in the set $\{4,5,6,7,8,9\}$, which is a strict superset of the initial set $\{4,6,9\}$.
This example shows that abstraction can result in imprecision. 

To reason with abstract values, all operations on concrete values must be ``lifted'' to operate on values in the abstract domain.
For an operation $\otimes$ on concrete-domain values,
we use $\abs{\otimes}$ to refer to its respective abstract counterpart.
For example, consider lifting addition ($+$) to the abstract addition operation ($+^\sharp$) to operate on abstract values in the interval domain:
the abstract transformer $+^\sharp$ is $[l_1, r_1] +^\sharp [l_2, r_2] = [l_1 + l_2, r_1 + r_2]$.
Consider two intervals $[5, 6]$ and $[10, 20]$;
their sum is $[5,6] +^\sharp [10,20] = [15, 26]$.
While the lifting of $+$ is straightforward, many simple operations, such as the absolute-value operation,
(\texttt{abs()}) have a non-trivial abstract transformer~\cite{amurth}: 
\begin{align}
  \mathtt{abs}^\sharp([\mathtt{l, r}]) = [\mathtt{max(max(0, l), -r), max(-l, r)}]. \label{eq:absIntervalBest}
\end{align}

%
As discussed in \Cref{sec:intro}, Kalita et al.\ introduced the problem of \emph{transformer synthesis modulo a domain-specific language (DSL)} for a single abstract domain \cite{amurth}.
Their method synthesizes an abstract transformer for an operation $\textit{op}$ over an abstract domain $\AbsDomain$, \emph{expressed in DSL $\Lang$}---what they call
an ``$\Lang$-transformer (for $\textit{op}$ over $\AbsDomain$).''
Their algorithm is guaranteed to return a sound and maximally precise $\Lang$-transformer.
As there may be other $\Lang$-transformers that are incomparable to the one obtained by the algorithm, which is why one says that the algorithm creates ``\emph{a} best $\Lang$-transformer.''

\subsection{Product Domains}
\label{sec:back:prod_dom}
It is possible to create \textit{product domains} that maintain information from multiple abstract domains, thereby improving the overall precision of the analysis.

\paragraph{Odd intervals and even intervals.}
We introduce two simple domains solely for illustrative purposes.
Each value in the \textit{odd-interval} abstract domain is a pair $[l_o, r_o]$. The domain constraints for the odd-interval domain enforces that, for abstract value $[l_o, r_o]$, $l_o$ and $r_o$ are \textit{odd} numbers, and $[l_o, r_o]$ abstracts a set of concrete values $S$ if and only if $l_o$ is less than or equal to all numbers in $S$, and $r_o$ is greater than or equal to all numbers in $S$.
More formally, for a non-empty set $S$,
\[
\begin{array}{r@{\hspace{0.75ex}}c@{\hspace{0.75ex}}l}
  \alpha(S)      & = &
     \left[\begin{array}{l}
        \inf(S) = -\infty \,?\, {-\infty} : (\textit{isOdd}(\textit{min}(S)) \,?\, \textit{min}(S) : (\textit{min}(S)\ -\ 1)), \\
        \sup(S) = \infty \,?\, \infty : (\textit{isOdd}(\textit{max}(S)) \,?\, \textit{max}(S) : (\textit{max}(S)\ +\ 1))
     \end{array}\right]
  \\
  \gamma([l, r]) & = & \{x~\in \mathbb{N} \mid l \leq x \leq r\}
\end{array}
\]

\begin{figure}[t]
\centering
\includegraphics[scale=.4]{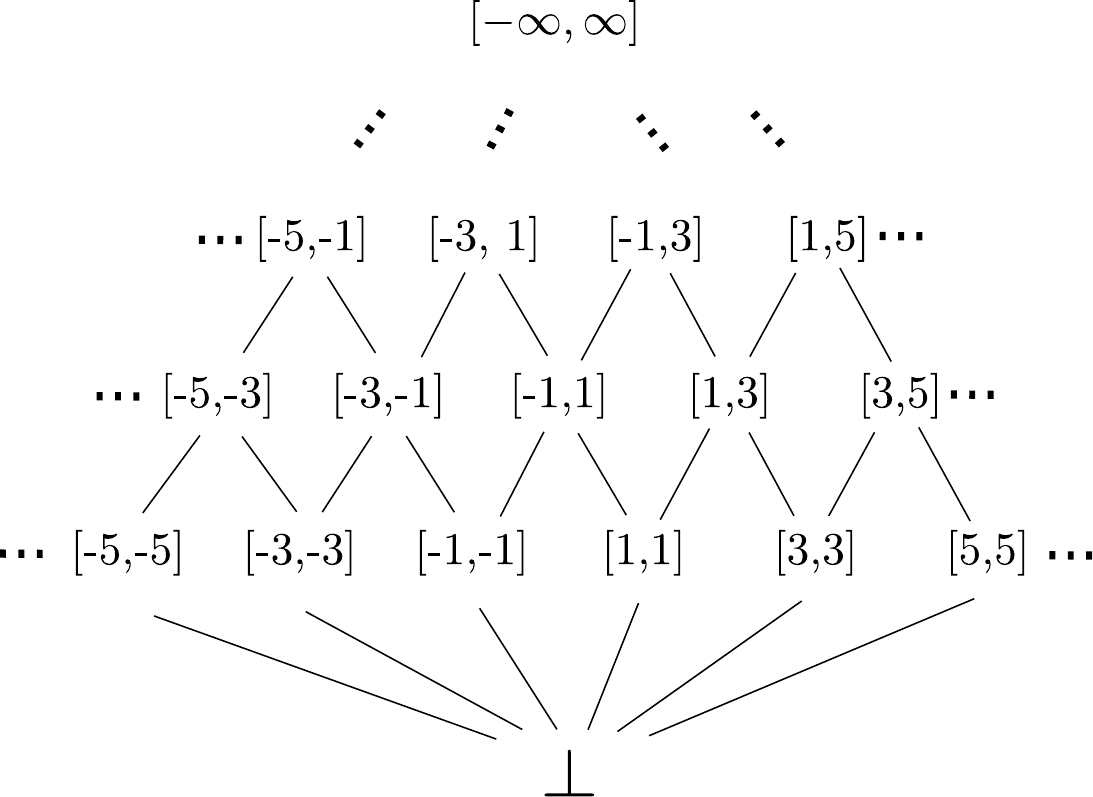}
\caption{Lattice for the odd-interval domain 
\label{fig:oddLattice}}
\end{figure}

\Cref{fig:oddLattice} shows the lattice for the odd-interval domain. 
The \textit{even-interval} domain is similar, except
that in an abstract value $[l_e, r_e]$, (finite) limits $l_e, r_e$ are now
required to be even.

\paragraph{Direct Products.}
A simple methodology for creating product domains is \textit{direct products}, where the product domain \textit{independently} applies the respective domain transformers to the abstract values from the component domains. For example, the direct-product transformer for the product domain of even-interval and odd-interval domains for the \textit{increment} operator is
\[
  \absd{inc}_{O\times E}(\langle a_o, a_e\rangle) = \langle \abs{inc}_O(a_o), \abs{inc}_E(a_e) \rangle.
\]

Note that the direct-product transformer can be computed merely by accumulating the results from the transformers of the component domains.

For example, \Cref{eq:incDirect} shows the transformer for the increment operation for the direct product of the odd-interval and even-interval domains.
\begin{align}
	\mathtt{inc^{\sharp D}(\langle o, e\rangle)} &= \langle \mathtt{[\colorbox{red!20}{\phantom{2}o.l\phantom{2}, o.r+2}]}, \mathtt{[\colorbox{blue!20}{\phantom{2}e.l\phantom{2}, e.r+2}]} \rangle\label{eq:incDirect}
 \end{align}

The transformer for the odd-interval domain is highlighted in red, and the
transformer for the even-interval domain is highlighted in blue.
The first interval $[\colorbox{red!20}{\phantom{2}o.l\phantom{2}, o.r+2}]$ is for the odd-interval domain, and the second one is for the even-interval domain.

Note that the abstract transformer given in \Cref{eq:incDirect} may not result in an answer that is as precise as the domain is capable of representing.
For instance, the most precise abstract value in the direct-product domain for the set $\{5\}$ is $\langle [5,5], [4,6] \rangle$.
Via \Cref{eq:incDirect},
\[
  \mathtt{inc^{\sharp D}(\langle [5,5], [4,6] \rangle)} = \langle [5,7], [4,8] \rangle,
\]
which represents the concrete set $\{ 5, 6, 7\}$.
While this answer is conservative, the domain is capable of representing the set $\alpha(\{ \mathtt{inc}(5) \})$ $= \alpha(\{6\})$ $= \langle [5,7], [6,6] \rangle$.
This example shows that the application of an abstract transformer can lead to loss in precision, but the result will be overapproximated.
 
 \paragraph{Reduced Products.}
An alternative is to work with a \textit{reduced-product} domain, which is similar to a direct-product domain, except that a \textit{reduction operator}, denoted by $\sigma$, is used to reduce the abstract value for each component domain to the \textit{smallest} possible abstract value in the respective domain that is consistent with the paired abstract value's concretization.
More precisely, suppose that $\alpha_1$ ($\gamma_1$) and $\alpha_2$ ($\gamma_2$) are the abstraction (concretization) functions for the respective domains of two abstract values $a_1$ and $a_2$.
The concretization of the pair $\langle a_1, a_2 \rangle$ is defined as follows: $\gamma(\langle a_1, a_2 \rangle) \eqdef \gamma_1(a_1) \cap \gamma_2(a_2)$.
If $c = \gamma(\langle a_1, a_2 \rangle)$, then $\sigma(\langle a_1, a_2 \rangle) = \langle \alpha_1(c), \alpha_2(c) \rangle$.
For instance, for the odd-interval/even-interval reduced-product domain, $\sigma(\b{o, e})=\b{[\textit{max}(o.l,\ e.l-1),\ \textit{min}(o.r,\ e.r+1)], [\textit{max}(o.l-1,\ e.l),\ \textit{min}(o.r+1,\ e.r)]}$. For example, $\sigma(\b{[3, 9], [-2, 6]})= \b{[3, 7], [2, 6]}$.\footnote{
  We assume that component arithmetic is extended to cover $-\infty$ and $\infty$---e.g., $-\infty - 1 = -\infty$, etc.
}

Reduced-product domains can lead to answers that are more precise than with direct-product domains, but the definitions of abstract transformers can be tricky.
To maximize precision with a reduced-product domain, one needs to use abstract transformers that create their answers as some function of all the abstract values of the individual domains participating in the product.
For example, \Cref{eq:incRed} shows the reduced-product transformer for the increment operation for the reduced product of the odd-interval domain and even-interval domain:\footnote{  We assume that the reduction operator $\sigma$ has always been applied before the transformer in \Cref{eq:incRed} is called.
}
\begin{align}
	\mathtt{inc^{\sharp R}(\langle o, e\rangle)} &= \langle\mathtt{[\colorbox{red!20}{e.l + 1, e.r+1}]}, \mathtt{[\colorbox{blue!20}{o.l + 1, o.r+1}]}\rangle\label{eq:incRed}
\end{align}

Consider the first interval (corresponding to the odd interval of the reduced product) in \Cref{eq:incRed}, $[\colorbox{red!20}{e.l+1, e.r+1}]$. It is quite interesting that
it obtains improved precision by using the parameters from the even-interval component for both the lower and upper limits of the odd-interval component of the answer.
Similarly, the parameters from the odd-interval component are used for both the lower and upper limits of the even-interval component of the answer.

\begin{figure}
\centering
\begin{subfigure}[b]{.45\linewidth}
\centering
\begin{tikzpicture}
\node (a1) {$\langle a_1$};
\node[right of=a1] (a2) {$a_2$};
\node[right of=a2] (d1) {$\ldots$};
\node[right of=d1] (an) {$a_n\rangle$};

\node[below of=a1] (f1) {$\langle \absd{f}_{1}$};
\node[below of=a2] (f2) {$\absd{f}_{2}$};
\node[below of=d1] (d2) {$\ldots$};
\node[below of=an] (fn) {$\absd{f}_{n}\rangle$};

\node[below of=f1] (ap1) {$\langle a'_1$};
\node[right of=ap1] (ap2) {$a'_2$};
\node[right of=ap2] (d3) {$\ldots$};
\node[right of=d3] (apn) {$a'_n\rangle$};

\draw[->] (a1) -- (f1);
\draw[->] (a2) -- (f2);
\draw[->] (an) -- (fn);
\draw[->] (f1) -- (ap1);
\draw[->] (f2) -- (ap2);
\draw[->] (fn) -- (apn);
\end{tikzpicture}
\caption{Direct-product transformers}
\end{subfigure}
\hspace{5mm}
\begin{subfigure}[b]{.45\linewidth}
\centering
\begin{tikzpicture}
\node (a1) {$\langle a_1$};
\node[right of=a1] (a2) {$a_2$};
\node[right of=a2] (d1) {$\ldots$};
\node[right of=d1] (an) {$a_n\rangle$};

\node[below of=a1] (f1) {$\langle \absr{f}_{1}$};
\node[below of=a2] (f2) {$\absr{f}_{2}$};
\node[below of=d1] (d2) {$\ldots$};
\node[below of=an] (fn) {$\absr{f}_{n}\rangle$};

\node[below of=f1] (ap1) {$\langle a'_1$};
\node[right of=ap1] (ap2) {$a'_2$};
\node[right of=ap2] (d3) {$\ldots$};
\node[right of=d3] (apn) {$a'_n\rangle$};

\draw[->] (a1) -- (f1);
\draw[->] (a2) -- (f2);
\draw[->] (an) -- (fn);
\draw[->] (f1) -- (ap1);
\draw[->] (f2) -- (ap2);
\draw[->] (fn) -- (apn);

\draw[->] (a2) -- (f1);
\draw[->] (an) -- (f1);
\draw[->] (a1) -- (f2);
\draw[->] (an) -- (f2);
\draw[->] (a1) -- (fn);
\draw[->] (a2) -- (fn);

\end{tikzpicture}
\caption{Reduced-product transformers}
\end{subfigure}
\caption{Illustrations of working of direct and reduced-product transformers\label{fig:illuDirRed}}
\end{figure}
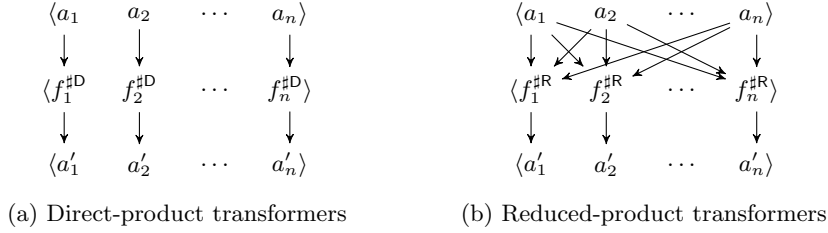

\Cref{fig:illuDirRed} illustrates the working of the direct and reduced-product transformers.
Note that 
$
  \mathtt{inc^{\sharp R}(\langle [5,5], [4,6] \rangle)} = \langle [5,7], [6,6] \rangle,
$
which represents the concrete set $\{ 6 \}$. Recall, the direct-product transformer produced a less precise concrete set ($\{5,6,7\}$) for the same operation. 

\smallskip
\noindent
\fbox{\parbox{.98\linewidth}{
For a product domain $A: A_1 \times A_2 \times \dots \times A_n$, we refer to $A_i$ as the \textit{component domains}. We denote reduced product abstract transformers as $\absr{f}:\b{\dfuns}$, where we refer to $\absr{f}_i$ as the \textit{component transformers} of $\absr{f}$. We denote direct product abstract transformers as $\absd{f}:\b{\absd{f}_1, \absd{f}_2, \dots, \absd{f}_n}$, where we refer to $\absd{f}_i$ as the \textit{component transformers} of $\absd{f}$. 
}
}

\subsection{Discussion}
Consider the increment transformer for the direct product of odd and even intervals.
With the initial abstract value $\langle[1,5], [2,6]\rangle$, the direct-product transformer (\Cref{eq:incDirect}) would yield $\langle[1,7], [2,8]\rangle$, which represents the concrete set $\{2,3,4,5,6,7\}$, which is more precise than both the even-interval and odd-interval abstract values (due to the absence of the numbers $1$ and $8$).
This example illustrates that a product domain can yield improved precision compared to the component domains.

\begin{figure}[t]
\centering
\begin{minipage}{.6\columnwidth}
\begin{minted}[autogobble, fontsize=\small, tabsize=2, escapeinside=||, breaklines]{C}
// |$\mathtt{a}_0^{\sharp D}: \langle[1,5], [2,6]\rangle = \{2,3,4, 5\}$\labelline{code:diffinitD}|
// |$\mathtt{a}_0^{\sharp R}: \langle[1,5], [2,6]\rangle = \{2,3,4,5\}$\labelline{code:diffinitR}|
|{\color{black}a++;}| 

// |$\mathtt{a}_1^{\sharp D}: \langle[1,7], [2,8]\rangle = \{2,3,4,5,6,7\}$|
// |$\mathtt{a}_1^{\sharp R}: \langle[3,7], [2,6]\rangle = \{3,4,5,6\}$|
|{\color{black}a++;}|

// |$\mathtt{a}_2^{\sharp D}: \langle[1,9], [2,10]\rangle = \{2,3,4,5,6,7,8,9\}$|
// |$\mathtt{a}_2^{\sharp R}: \langle[3,7], [4,8]\rangle = \{4,5,6,7\}$|
|{\color{black}a++;}|

// |$\mathtt{a}_3^{\sharp D}: \langle[1,11], [2,12]\rangle = \{2,3,4,5,6,7,8,9,10,11\}$\labelline{code:diffendD}|
// |$\mathtt{a}_3^{\sharp R}: \langle[5,9], [4,8]\rangle = \{5,6,7,8\}$ \labelline{code:diffendR}|
\end{minted}
\end{minipage}
\caption{An example to show precision in both direct and reduced product\label{fig:diffRD}}
\end{figure}

The reader may wonder whether the additional complications involved in defining reduced-product abstract transformers (and proving them correct) is worth it.
\Cref{fig:diffRD} shows three applications of the increment function, and illustrates how imprecision can snowball.
For a single step, the concretized output set of resulting values from the reduced-product transformer is more precise than what we obtain using the direct-product transformer.
The difference in precision is magnified by subsequent applications of increment, and overall significantly better precision for the sequence of statements is obtained using the reduced-product transformers. This example
shows that reduced products can result in significantly improved precision than what direct products can provide.

However, constructing a provably sound and most-precise reduced-produce transformer is challenging. 
This article proposes a practical solution for automatically constructing such transformers.

\section{Overview}
\label{sec:overview}

\subsection{Problem Statement}
\label{Se:ProblemStatement}

In this paper, we consider \emph{transformer synthesis for multiple abstract domains, modulo multiple DSLs}.
The problem is defined as follows: 

\smallskip
\noindent
\fbox{\parbox{.98\linewidth}{
Given a concrete domain (C), a set of abstract domains, ${A_1, A_2, \dots, A_n}$, a concrete operation $f$, and domain-specific languages (DSLs) $\Lang_1, \Lang_2, \ldots, \Lang_n$, the goal is to synthesize a sound and most precise \textit{reduced} abstract transformer $\absr{f}:\langle \absr{f}_{1}, \absr{f}_{2}, \dots, \absr{f}_{n} \rangle$ for the product domain $\mathcal{D}: A_1 \times A_2 \times \dots \times A_n$, where the $i^{\textit{th}}$ component of $\absr{f}$ is expressed in DSL $\Lang_i$. We use $\Lang$ to denote this tuple of languages $\b{\Lang_1, \Lang_2, \dots, \Lang_n}$ for the component abstract domains.
}
}

\smallskip
\noindent
Thus, our work addresses a different problem from prior work on reduced products:
the goal is to create abstract transformers for a reduced-product domain, but the problem is parameterized by a collection of DSLs $\Lang_1, \Lang_2, \ldots, \Lang_n$
in which the component abstract transformers are to be expressed.
Our algorithm attempts to return an abstract transformer---expressed using $\Lang_1, \Lang_2, \ldots, \Lang_n$ for the respective components---that is one of the collection of incomparable most-precise (``best'') abstract transformers expressible with those languages.
We assume that the concrete operation is provided symbolically as a logical formula. If it is provided as a 
program\footnote{The concrete operation can be expressed as a loop-free program, or a program with bounded loops.}, standard encodings are available to encode it in logic~\cite{POPL:flanagan01,CACM:hoare69}.
Each DSL $\Lang_i$ is provided as a context-free grammar $\mathcal{G}_i$, along with semantics that is specified on a production-by-production basis.
For each abstract domain $A_i$, we require the following:
\begin{itemize}
  \item
    \textit{A complete lattice over abstract values} $(A_i, \sqsubseteq_i, \bot_i)$,
    where $A_i$ is the set of abstract values in the abstract domain, $\sqsubseteq_i$ is the (partial) ordering relation amongst the abstract values and $\bot_i$ is the least element in $A_i$. 
  \item
    \textit{A Galois connection that relates the abstract and concrete domains.}
    A Galois connection is defined by monotonic functions
    $\alpha_i: \mathcal{P}(C) \rightarrow A$ and $\gamma_i: A_i \rightarrow \mathcal{P}(C)$,
    for which for all $a \in A_i$, $c \in \mathcal{P}(C)$
    \[
      \alpha(c) \sqsubseteq a \Leftrightarrow c \subseteq \gamma(a).
    \]
    where $\mathcal{P}$ denotes the powerset of a provided set.
\end{itemize}

\noindent

In \Cref{sec:product-overview}, we propose weaker notions of precision, motivated by the desire to create an algorithm that works in practice.

\subsection{Challenges}
\label{Se:Challenges}

\begin{figure}[t]
    \centering
    \begin{subfigure}[b]{.45\linewidth}
        \centering
        \includegraphics[scale=.7]{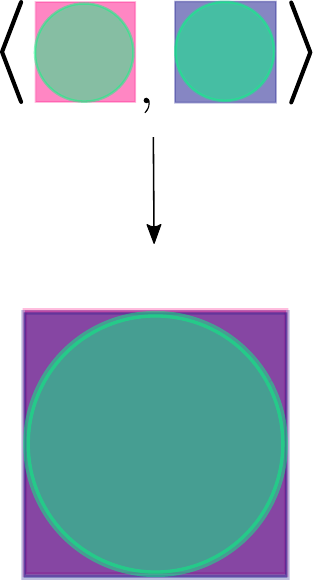}
        \caption{Independent Synthesis\label{fig:independent}}
    \end{subfigure}
    \hfill
    \begin{subfigure}[b]{.45\linewidth}
        \centering
        \includegraphics[scale=.7]{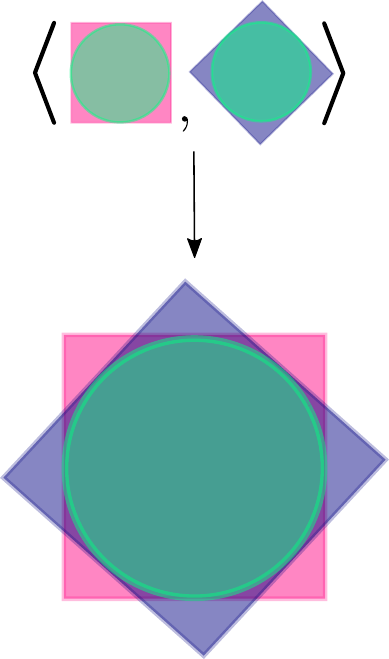}
        \caption{Dependent Synthesis\label{fig:dependent}}
    \end{subfigure}
    \caption{An illustration to show the benefit of synthesizing with dependency on other domains over synthesis of transformer for each domain independently\label{fig:dependent_independent}}
\end{figure}

There are really two separate challenges that must be addressed to synthesize a best, multi-domain, abstract transformer modulo a collection of DSLs:
\begin{enumerate}
  \item
    \label{It:DomainOfDiscourse}
    \textit{Enlarging the domains of discourse.}
    As illustrated in \Cref{eq:incRed}, the computation performed by the abstract transformer for one component domain may need to access information from one or more of the other component domains.
    Thus, compared with what is needed to synthesize the abstract transformer for a component domain $D_i$ in a direct-product construction (e.g., by invoking the vanilla \amurth algorithm for each component independently), each DSL $\Lang_i$ may need to be enlarged to allow expressing computations using information from (the representations of) the abstract values in each other component domain.
  \item
    \label{It:BestPlusBest}
    $\textit{best}_i + \textit{best}_k \neq \textit{best}$.
    As discussed in \Cref{sec:intro}, because we are dealing with $\Lang_i$-transformers for various DSLs $\Lang_i$, there is no assumption of there being a single best $\Lang_i$-transformer.
    On the contrary, there can be a \emph{collection} of incomparable best $\Lang_i$-transformers (one of the challenges addressed in the \amurth paper \cite{amurth}).
    \Cref{fig:dependent_independent} demonstrates that even without extending the DSLs so that $\Lang_i$ can access components of other domains $D_k$ (i.e., \Cref{It:DomainOfDiscourse}), not every combination of a best $\Lang_i$-transformer and a best $\Lang_k$-transformer gives you a best ($\Lang_i$ x $\Lang_k$)-transformer.
    (See the discussion below.)
\end{enumerate}

\Cref{It:DomainOfDiscourse} is what one might expect from the notion of reduced product: there needs to be some means for communicating information among domains, so in the context of synthesis modulo a collection of DSLs, an obvious mechanism is to enlarge the domain of discourse of each DSL.

\Cref{It:BestPlusBest} is a separate consequence of taking the synthesis-modulo-DSL-$\Lang$ problem from \amurth and extending it to the multi-domain, multi-DSL setting.
It does not have an analogue in standard treatments of reduced product, so in that sense, \Cref{It:BestPlusBest} has more surprise value than \Cref{It:DomainOfDiscourse}.

\Cref{It:BestPlusBest} can be illustrated as follows:
\Cref{fig:dependent_independent} shows two ways of abstracting a circular region of reachable states (green background) by square abstractions (magenta and blue).
Let us assume that we use a product domain of two square abstractions for improved precision, but that each DSL can only create a transformer that produces (i) a square that is aligned with the $x$ and $y$ axes, or (ii) a square that is aligned at $45^\circ$ to the axes.\footnote{
  Strictly speaking, \Cref{fig:dependent_independent} illustrates just the abstract values produced as the post-transformation abstract state, rather than the abstract transformers \emph{per se}.
}

\Cref{fig:independent} shows a case that is possible when the transformer for each domain is synthesized independently.
With no knowledge of what the other transformer produces, each synthesis run finds \emph{a} best transformer for each domain---and in this case ends up with two transformers that yield the same square.
Consequently, the product transformer also yields the same square---i.e., it has no better precision than either of the component transformers.

Because for each of the domains $D_1, D_2, \ldots$ and corresponding DSLs $\Lang_1, \Lang_2, \ldots$, one can only obtain \underline{\textit{a}} best $\Lang_i$-transformer, among all the combinations of best $\Lang_1$-transformers and best $\Lang_2$-transformers there can be better and worse combinations.
For instance, the two purple squares in \Cref{fig:dependent_independent} are results produced by two different best $\Lang_2$-transformers.
However, the purple ($\Lang_2$) square that is rotated by $45^\circ$ in \Cref{fig:dependent}, when combined ($\cap$) with the axis-aligned magenta ($\Lang_1$) square, produces a strict subset of the result shown in \Cref{fig:independent}.
The result shown in \Cref{fig:dependent} is what a best $(\Lang_1 \times \Lang_2)$-transformer should produce, whereas a transformer that produces the result shown in \Cref{fig:independent} would not be a best $(\Lang_1 \times \Lang_2)$-transformer.

This example has elucidated an important property of the algorithm to synthesize a best $\Lang$-transformer in the multiple-abstract-domain, multiple-DSL setting, namely,

\smallskip
\noindent
\fbox{\parbox{.98\linewidth}{
Synthesis of each domain's transformer must be \textit{conditioned on the other domain transformers} such that the overall precision of the reduced transformer is improved.
}}

\subsection{Automatically Synthesizing Reduced Abstract Transformers}

One approach to the synthesis of abstract transformers is to explicitly construct the product domain and attempt to synthesize the transformers for it.
Because there exists prior work, \amurth~\cite{amurth}, that is capable of synthesizing a best $\Lang$-transformer for a given domain, \amurth applied to the product domain can synthesize a best multi-domain, multi-DSL transformer.
Using a given DSL $\Lang$,
\amurth runs two counterexample-guided-inductive-synthesis loops for soundness and precision to yield a provably sound and most precise $\Lang$-transformer.
However, this method does not scale because it requires the component transformers for each of the domains to be synthesized simultaneously:
the transformer for the addition operation that our tool, \tool, synthesizes in about half-an-hour cannot be synthesized by \amurth in over 10 hours.

As opposed to synthesizing the transformers for all domains at the same time, \tool attempts to synthesize, one by one, the abstract transformer for each component domain in the product domain while keeping the transformers for all other component domains fixed. 

We begin our discussion by understanding the synthesis algorithm for a single domain from \amurth \cite{amurth} (\S\ref{sec:single-overview});
we then discuss why that algorithm does not work for reduced-product domains, and discuss our primary contribution---a novel algorithm for reduced-product domains (\S\ref{sec:product-overview}).

\begin{figure}
\centering
\includegraphics[scale=.45]{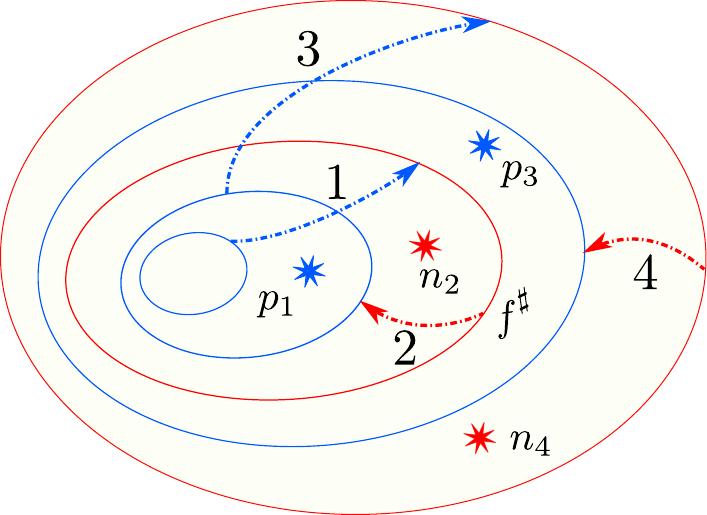}
\caption{Illustration of shrinking and expanding of transformer $\abs{f}$ \label{fig:expand-shrink}}
\end{figure}

\subsection{Automated $\Lang$-Transformer Synthesis for a Single Domain}
\label{sec:single-overview}

For simplicity, let us consider abstract transformers of arity one, i.e., functions, $\abs{f}: A \rightarrow A$. However, because we only have access to the concrete function, $f: C \rightarrow C$, we use examples $\langle a, c'\rangle$, where $a \in A$ is an abstract input, and $c' \in \{f(c) \mid c \in \gamma(a) \}$.
We may define an \textit{ideal} transformer by pointwise application of the concrete transformer, $f$:
\begin{align} 
\widehat{\abs{f}}(a) \equiv \alpha(\{ f(c) \mid c \in \gamma(a) \}) \label{eq:bestAbsTrans}
\end{align}

Because we are interested in synthesizing an \textit{executable} function that satisfies the syntactic constraints posed by the DSL with which we are working, we search for the \textit{best over-approximation} of $\widehat{\abs{f}}$ that can be expressed in $\Lang$.

\paragraph{Soundness.} We say that an $\Lang$-transformer $\abs{f} \in \Lang$ is sound iff 
\[ \forall a.\ \widehat{\abs{f}}(a) \sqsubseteq \abs{f}(a) \]

Hence, given a abstract transformer $\abs{f}$ and a concrete function $f$, a counterexample to soundness is a pair $\b{a, c'}$ that makes the following formula satisfiable:
\[
  \exists c \in \gamma(a).\  c' = f(c) \land c' \notin \gamma(\abs{f}(a))
\] 

The candidate transformer $\abs{f}$ must be \textit{expanded} to include such counterexamples;
i.e., we synthesize a new transformer that is consistent with all the (positive and negative) counterexamples generated so far, and also includes the positive counterexample just generated.
Hence, we refer to counterexamples to soundness as \textit{positive counterexamples}.

\paragraph{Precision.}

We say that an $\Lang$-transformer $\abs{f}$ is maximally precise if, for all sound $\Lang$-transformers $\abs{h}$ that are comparable to $\abs{f}$,
for all abstract inputs $a$, $\abs{f}(a) \sqsubseteq \abs{h}(a)$:
\[
  \forall \abs{h} \in \Lang.\ (\textit{isSound}(\abs{h}) \land \textit{comparable}(\abs{f}, \abs{h})) \implies (\forall a \in A.\ \abs{f}(a) \sqsubseteq \abs{h}(a)),
\]
where $\textit{comparable}(\abs{g_1}, \abs{g_2})$ returns \textsf{true} if $\forall a\in A.\ \abs{g_1}(a) \sqsubseteq  \abs{g_2}(a) \lor \abs{g_2}(a) \sqsubseteq \abs{g_1}(a)$.


Hence, a counterexample to precision requires a ``witness'' $\Lang$-transformer $h$ that is \textit{strictly} more precise than the candidate $\Lang$-transformer $\abs{f}$.
That is, $\b{a,c'}$ is a counterexample to precision iff 
\[
  \exists \abs{h} \in \Lang.\ 
        \textit{isSound}(\abs{h})
  \land \textit{comparable}(\abs{f}, \abs{h})
  \land \exists a \in A.\ \exists c' \in \gamma(\abs{f}(a)).\ c' \notin \gamma(\abs{h}(a)).
\]

The candidate $\Lang$-transformer $\abs{f}$ must be \textit{shrunk} to exclude such counterexamples, i.e., we synthesize a new $\Lang$-transformer that is consistent with all the (positive and negative) counterexamples generated so far, while also excluding the current negative counterexample ($c$) generated. Hence, we refer to counterexamples to precision as \textit{negative counterexamples}.

Note that the set of all abstract $\Lang$-transformers forms a partial order with respective to the precision relation, and hence there may be multiple incomparable abstract $\Lang$-transformers that are maximally precise. For example, for a function that always returns zero ($\lambda x.0$)
and a DSL $\Lang$ over intervals that is required to have one of its limits grounded at zero,
i.e., ($\{ \lambda I.[0, i], \lambda I. [-i, 0] \mid i \in \mathbb{N} \land i \neq 0\}$),
there exist two maximally precise but incomparable $\Lang$-transformers, i.e., $\lambda I.[-1, 0]$ and $\lambda I.[0, 1]$.
Note that both {$\b{[0,0], 1}$} and {$\b{[0,0], -1}$} are potential negative counterexamples ({$\b{[0,0], 1}$} is a counterexample to $\lambda I.[-1, 0]$ and {$\b{[0,0], -1}$} is a counterexample to $\lambda I.[0, 1]$). 
However, adding both counterexamples will make the synthesis problem unsatisfiable because both of the maximally precise $\Lang$-transformers would be disallowed.
Hence, counterexamples to precision can only be treated as \textit{soft} counterexamples---we attempt to satisfy most of them en route to synthesis of one of the maximally precise $\Lang$-transformers.

\paragraph{Algorithm for synthesizing $\Lang$-transformers.} Starting with any initial candidate $\Lang$-transformer (say, $\lambda a. \abs{\bot}$), we non-deterministically cycle through the two phases---the expansion phase (by generating a positive counterexample), and the shrinking phase (by the generation of a negative counterexample)---until no further counterexamples can be generated. The reader can check \amurth~\cite{amurth} for more details about synthesizing $\Lang$-transformers for a single domain. \Cref{fig:expand-shrink} illustrates the working of this algorithm. The red points are negative counterexamples, while the blue points denote positive counterexamples; the oval shapes denote a candidate abstract transformer $\abs{f}$. An example $\b{a,c'}$ is shown to lie within a transformer if $c' \in \gamma(\abs{f}(a))$. 


\subsection{Our Contribution: Automated $\Lang$-Transformer Synthesis for Reduced-Product Domains}
\label{sec:product-overview}

An $\Lang$-transformer for a reduced-product domain $\mathcal{D}=A_1 \times A_2 \times \dots \times A_n$ and a tuple of DSLs
$\Lang$ $= \langle \Lang_1, \Lang_2, \ldots, \Lang_n \rangle$
is a tuple of $\Lang_i$-transformers, one for each of the component domains: $\absr{f}: \b{\absr{f}_1, \absr{f}_2, \dots, \absr{f}_n}$. We denote the abstraction and concretization functions of each component domain $A_i$ by $\alpha_i$ and $\gamma_i$, respectively,
and the abstraction and concretization functions for $\mathcal{D}$ by $\alpha$ and $\gamma$.

The $\Lang$-transformer of the reduced-product domain, $\absr{f}$ must satisfy the following property with respect to the $\Lang_i$-transformers $\absr{f}_i$ for the component domains:
\[
  \absr{f}(a) = a' \implies \gamma(a') = \bigcap_{i=1}^n \gamma_i(\absr{f}_i(a))
\]

This property implies,
\begin{itemize}
	\item $ c' \in \gamma(\absr{f}(a)) \implies  \bigwedge_{i=1}^n c' \in \gamma_i(\absr{f}_i(a)) $
	\item $ c' \notin \gamma(\absr{f}(a)) \implies \bigvee_{i=1}^n c' \notin \gamma_i(\absr{f}_i(a)) $
\end{itemize}

That is, a concrete output value $c'$ is included in the output of a product transformer if it is included by \textit{all} the component domain transformers; on the other hand, a concrete output value $c'$ is excluded in the output of a product transformer if it is excluded by \textit{any} of the component domain transformers.

\begin{figure}
\centering
\begin{subfigure}[b]{.45\linewidth}
    \centering
    \includegraphics[scale=0.45]{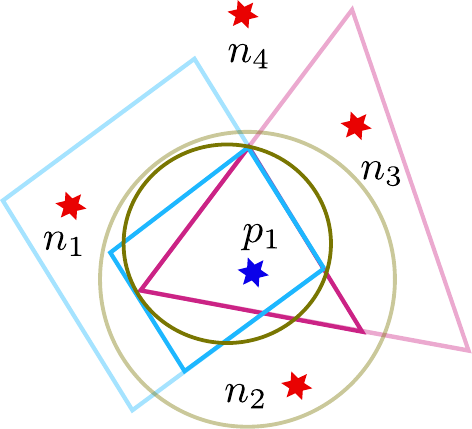}
    \caption{Illustration of different transformer results from three different abstract domains\label{fig:reduced_cirtrisq}}
\end{subfigure}
\hfill
\begin{subfigure}[b]{.45\linewidth}
    \centering
    \includegraphics[scale=0.45]{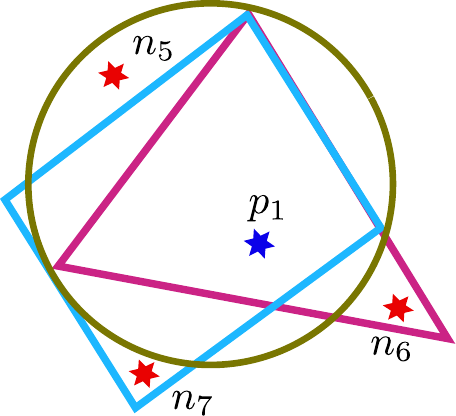}
    \caption{
    Enlarged version of \Cref{fig:reduced_cirtrisq}\label{fig:reduced_cirtrisq_zoom} showing some additional kinds of negative examples.
    }
\end{subfigure}
\caption{Positive and negative examples in reduced product domain\label{fig:posex_negex_prod}}
\end{figure}

For synthesizing reduced $\Lang_i$-transformers for a reduced-product domain $\mathcal{D}=A_1 \times \dots \times A_n$, we use examples of the form $\b{\b{a_1,\dots, a_n}, c'}$ for learning the reduced transformer, where $a_1\in A_1, a_2 \in A_2, \dots, a_n\in A_n$.
An example $\b{\b{a_1,\dots, a_n}, c'}$ is a positive example for $\absr{f}$ if $c' \in \gamma_i(\absr{f}_i(\b{a_1,\dots, a_n}))$ for each domain $A_i$.
On the other hand, an example $\b{\b{a_1,\dots,a_n},c'}$ is a negative example for $\absr{f}$, if there exists some domain $A_i$ such that $c'\notin \gamma_i(\absr{f}_i(\b{a_1,\dots, a_n}))$. 
These observations imply
\begin{itemize}
  \item{\it shared positive examples.}
    Positive examples must be maintained globally across all domains, allowing a positive counterexample $\b{\b{a_1,\dots, a_n}, c'}$ discovered while working on one domain transformer, to be automatically available as a positive example for all other domain transformers.
  \item {\it private negative examples.}
    Negative examples must be maintained privately for each of the component domains.
\end{itemize}

\Cref{fig:posex_negex_prod} shows positive and negative examples in product domains. Each shape corresponds to a component transformer for a reduced-product domain. The red points are negative examples, while the blue points refer to positive examples. We denote an example $\b{\b{\args}, c'}$ to lie inside a component transformer $\absr{f}_i$ if $c' \in \gamma_i(\absr{f}_i(\args))$. In contrast to single domain cases, the examples have a lot more variety: for example, $n_5$ is a negative example for $\absr{f}$ even though it is inside the component transformer $\absr{f}_{circle}$ (as it is outside the other component transformers). Note that the example $p_1$ is a positive example as it lies inside all the component transformers.

Let us now ``lift" the notions of soundness and precision to product domains, where $\widehat{\absr{f}}$ is the ideal transformer of the reduced-product domain.

\paragraph{Soundness.}
For a candidate $\Lang$-transformer for the reduced-product domain $\b{\absr{f}_1, \dots, \absr{f}_n}$ to be sound, the concretization of the post-abstract value must be overapproximated by all the component domain transformers.
\[
  \forall a.\ \bigwedge_{i=1}^n \gamma(\widehat{\absr{f}}(a)) \subseteq \gamma_i(\absr{f}_i(a)).
\]

\paragraph{Precision.}
A candidate reduced $\Lang$-transformer $\absr{f}=\b{\absr{f}_1, \absr{f}_2, \dots, \absr{f}_n}$ is precise if there does not exist a witness $\Lang$-transformer $\absr{h}=\{\absr{h}_1, \dots, \absr{h}_n\}$, $\absr{f}_i$ being the transformers corresponding to the component domains $A_i$, such that replacing each $\absr{f}_i$ by $\absr{h}_i$ leads to a more precise concretization set for the reduced-product post-state abstract value.
Hence, a candidate $\Lang$-transformer $\absr{f}$ is precise if the following formula is unsatisfiable:
\[
  \exists a.\ \exists \b{\absr{h}_1, \dots, \absr{h}_n}.\ \bigcap_{i=1}^n \gamma(\absr{h}_i(a)) {\subset} \bigcap_{i=1}^n \gamma_i(\absr{f}_i(a)) \land \bigwedge_{i=1}^n \textit{isSound}(\absr{h}_i).
\]

\paragraph{k-precision.}
The above formulation requires us to synthesize all the component transformers together, which does not scale well. A compromise is to limit this search for ``better'' transformers to subsets of all possible component transformers:

\[
  \exists a.\ \exists H.\ \bigcap_{i=1}^n \gamma(\abs{g}_i(a)) \subseteq \bigcap_{i=1}^n \gamma_i(\abs{f}_i(a)) \land \bigwedge_{i=1}^n \textit{isSound}(\absr{h}_i) \land |H| \leq k
\]

where, 
\begin{align}
\label{eq:ElementsOfH}
\abs{g}_i &= 
    \begin{cases}
        \abs{f}_i & \abs{h}_i \notin {H} \\
        \abs{h}_i & \abs{h}_i \in {H}
    \end{cases}
\end{align}

In particular, the simplest possible case is of 1-precision, where we only search if it is possible to find a single abstract transformer that can improve the precision of the resultant abstraction. The check of 1-precision can be simplified to $n$ different second-order queries, one for each component transformer.

We conjecture that $k$-precision {can be} weaker than $(k+1)$-precision: there may exist cases where changing $(k+1)$ component transformers simultaneously
may improve the precision of the resultant reduced transformer in a way that changing any $k$ component transformers cannot achieve. However, as of now, we could neither find a proof nor a counterexample to equivalence of $k$-precision and $(k+1)$-precision; in all our experiments, we only attempt to synthesize maximally 1-precise transformers, but all the synthesized reduced-product transformers were found to be maximally precise.
When $k$ equals the number of component domains, $k$-precision reduces to
checking precision
in the reduced-product domain. In the rest of the paper, we refer to ``best" $\Lang$-transformers as sound and maximally 1-precise transformers that are expressible in a language $\Lang: \b{\Lang_1, \Lang_2, \dots, \Lang_n}$.

\begin{figure}[t]
    \centering
    \includegraphics[scale=.25]{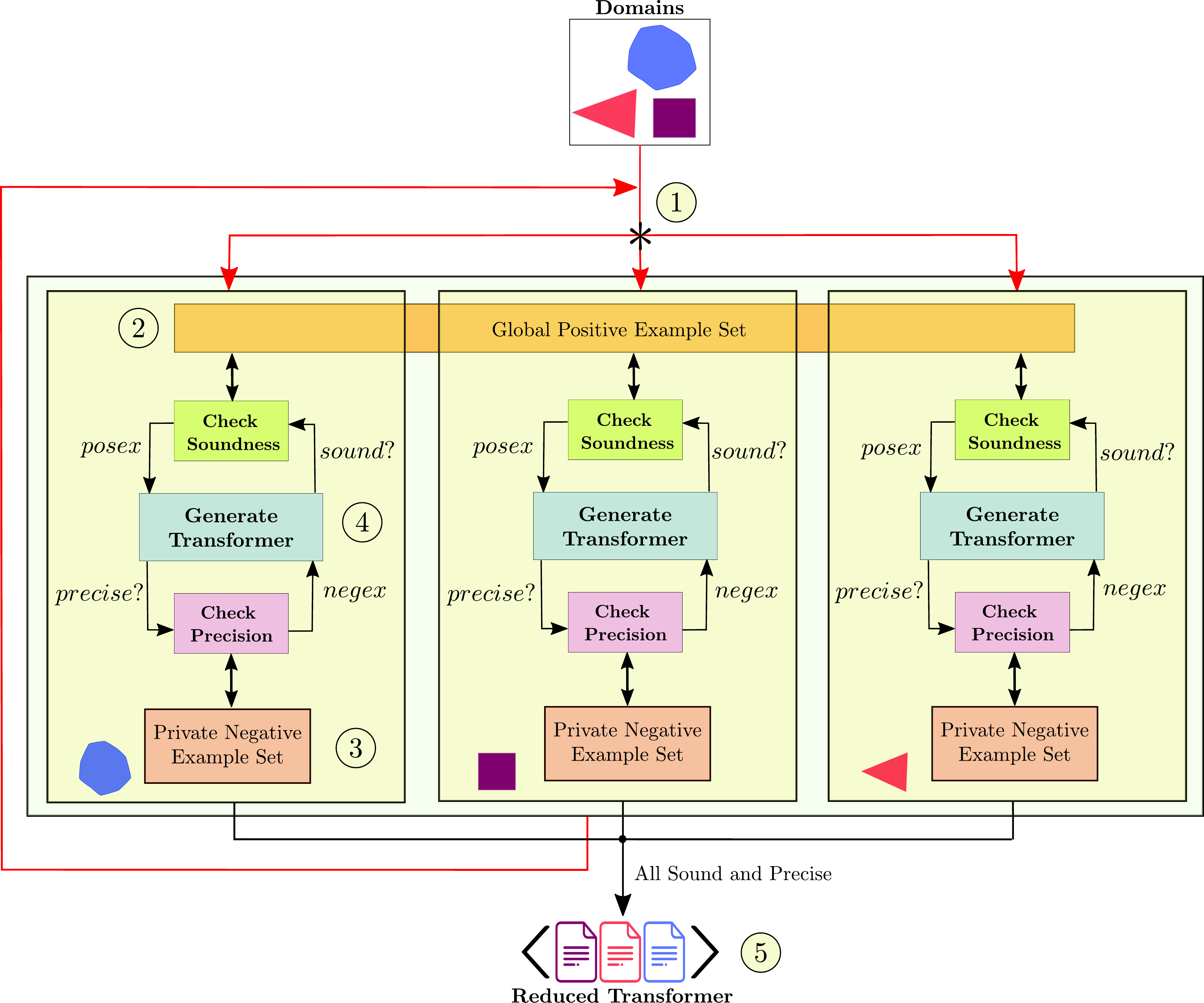}
    \caption{Overview of \tool \label{fig:algoFlowchart}}
\end{figure}

\paragraph{High-level algorithm for synthesizing reduced $\Lang$-transformers}

\Cref{fig:algoFlowchart} shows a high-level schematic of the algorithm in \tool.
The algorithm limits its search to 1-precision for scalability, i.e., for each of the domain $\Lang_i$-transformers, it independently searches for a better $\Lang_i$-transformer \textit{while fixing all other transformers}.
Our algorithm non-deterministically chooses one of the component domains and attempts to check if it is both sound and 1-precise \circled{1}. 

If a soundness counterexample is discovered in any domain, this positive example is shared with all other domains \circled{2}. If a 1-precision counterexample is discovered, it is added to the private set of negative examples for the respective domain \circled{3}. Then, a new domain $\Lang_i$-transformer is synthesized with respect to the augmented set of examples \circled{4}. Essentially, we run two counterexample-guided inductive synthesis (CEGIS) loops for each of the component domains in the product---one for soundness and the other for 1-precision. If no such soundness or 1-precision counterexamples are found, the algorithm goes back to non-deterministically choosing another domain \circled{1}. 
When all the domain $\Lang_i$-transformers are validated as sound and 1-precise, the reduced-product domain transformer is returned \circled{5}.

\paragraph{Running Example.} Let us now illustrate the high-level working of our algorithm with an example
and a possible run of the algorithm:

We use the increment operation (\texttt{++}) to explain each step of \tool with the odd and even interval domains as the component domains.
The following is the DSL
$\Lang_i$ used for both component domains during
the synthesis procedure.
\begin{align}
  \label{eq:intervalDSL}
  \textit{F}^\sharp ::=\ & \lambda \langle\texttt{o, e}\rangle . \langle[E, E], [E, E]\rangle \nonumber \\
                     E ::=\ & \texttt{o.l} \mid \texttt{o.r} \mid  \texttt{e.l} \mid \texttt{e.r} \mid 0 \mid 1 \mid {-}E \mid E + E \mid E - E \mid \nonumber \\
                            & \texttt{min}(E, E) \mid \texttt{max}(E, E) \mid {+}\infty \mid {-}\infty 
\end{align}
In the first phase, say \tool non-deterministically chooses the odd-interval component domain (\circled{1}), and generates a positive example $\langle\langle[27,29], [28, 30]\rangle, 30\rangle$.
This positive example will be added to the global set of positive examples \circled{2},
and a new $\Lang$-transformer is synthesized for the odd-interval component domain.
Subsequently, the 1-precision check for the odd-interval domain may identify $\langle\langle [-27, -25], [-28, -26]  \rangle, -22\rangle$ as a negative example,
which will be added to the private set negative examples for the odd-interval component domain \circled{3}. With these examples, \tool may synthesize
the following component $\Lang$-transformer for the odd-interval domain:
\begin{align}
    \repo{inc}{o,e}_{O} = [\mathtt{min(e.l + 1,  o.r), o.r+2}]
\end{align}
Because this $\Lang$-transformer is not sound or 1-precise
(for the odd-interval domain), the search carried out for the odd-interval domain's $\Lang$-transformer 
will go through a few rounds of soundness and 1-precision checks before emitting the following component transformer for the odd-interval component domain:
\begin{align}
  \label{eq:OddIntervalTransformerCorrect}
    \repo{inc}{o,e}_{O} = [\mathtt{e.l + 1, e.r + 1}]
\end{align}

Because this $\Lang$-transformer is both sound and 1-precise for the odd-interval domain, \tool
will break out of the CEGIS loops for the odd-interval domain, and return to \circled{1} to non-deterministically choose a new domain that is still not sound and 1-precise.
In this case, it will end up selecting the even-interval domain.
Similar to the odd domain, the search carried out for the even domain's $\Lang$-transformer also goes through a few iterations of soundness and 1-precision checks to finally synthesize the following component $\Lang$-transformer for the even-interval domain. 
\begin{align}
    \repo{inc}{o,e}_{E} = [\mathtt{o.l + 1, o.r + 1}]
\end{align}

Because the above $\Lang$-transformer is sound and 1-precise for the even-interval domain, \tool will
return back to \circled{1} to non-deterministically select a domain that is not yet sound and 1-precise.

However, because the $\Lang$-transformer from \Cref{eq:OddIntervalTransformerCorrect}, synthesized as the odd-interval domain's transformer, was both sound and 1-precise, and the sound/1-precise status of the even-interval domain's $\Lang$-transformer continues to hold, \tool returns \Cref{eq:incRed} as the final sound and 1-precise
reduced transformer for the reduced-product domain of the odd-interval and even-interval domains \circled{5}.  

\section{Algorithm}
\label{sec:algorithm}

	


Our algorithm follows a counterexample-guided inductive synthesis (CEGIS) strategy to synthesize an $\Lang$-transformer for reduced-product domains.
As in \Cref{sec:product-overview}, we start by describing the general algorithm (which is not scalable), before describing the specialization that supports 1-precision.
The algorithm generates positive examples (counterexamples to soundness) and negative examples (counterexamples to precision), accumulating them in a set of positive examples, $\pex$, and a set of negative examples, $\nex$, respectively. The algorithm converges to a sound and precise $\Lang$-transformer when neither a positive nor a negative example can be generated.

\subsection{Checking Soundness}

\paragraph{Positive Examples.} We say that a candidate $\Lang$-transformer $\b{\dfuns}$ for a reduced-product domain $\D = A_1 \times \dots \times A_n$ satisfies a positive example $\b{\b{\args}, c'}$ 
if $\b{\b{\args}, c'}$ is satisfied by each of the $\absr{f}_i$:

    \[\bigwedge_{A_i \in \D} c' \in \gamma(\absr{f}_i(\args))\]

We are now in a position to describe the complete soundness check: a given transformer $\b{\dfuns}$ is not sound on a set of examples in $\pex$ if there exists a counterexample $\b{\b{\args}, c'}$ such that,
 
\begin{align}
\exists A_k \in \D.\ \exists c \in \mathcal{C}.\ \big(\bigwedge_{i=1}^{n} c \in \gamma_i(a_i) \big)\land c' = f(c) \land \big( c' \notin  \gamma_k(\absr{f}_{k}(a_1, \dots, a_{n}))  \label{eq:ReducedCheckSoundness1}
\end{align}

The above can be realized as independent checks for each of the component domain $\Lang_i$-transformers:
\begin{align}
\exists c \in \mathcal{C}.\ \big(\bigwedge_{i=1}^{n} c \in \gamma_i(a_i) \big)\land c' = f(c) \land \big( c' \notin  \gamma_k(\absr{f}_{k}(a_1, \dots, a_{n}))  \label{eq:ReducedCheckSoundness}
\end{align}

We define the following interface for \textsc{CheckSoundness} that performs the above check (\Cref{eq:ReducedCheckSoundness}):
\begin{align}
\textsc{CheckSoundness}(& \absr{f}_{k}, f) = \\
&\begin{cases} \nonumber
\textit{False}, \langle\langle a_1, \ldots, a_{n}\rangle, c' \rangle  & \textrm{if \Cref{eq:ReducedCheckSoundness} is SAT}\\
\textit{True}, \_ & \text{otherwise}
\end{cases}
\end{align}

\subsection{Checking Precision}

\paragraph{Negative Examples.} We say that a candidate reduced-product $\Lang$-transformer $\absr{f}: \b{\dfuns}$ for a reduced-product domain $\D = A_1 \times \dots \times A_n$ satisfies a negative example $\b{\b{a_1,\dots, a_n}, c'}$,
if $\b{\b{a_1,\dots, a_n}, c'}$ fails to hold for at least one of the $\absr{f}_i$:

    \[\exists A_i \in \D.\ c' \notin \gamma_{\pankaj{i}}(\absr{f}_i(\args))\]

We extend the definition of positive and negative examples from satisfying a single example to a set of examples in $\pex$ or $\nex$. First, let us define predicates $\textit{satI}^+$ ($\textit{satI}^-$) to capture the condition that a domain transformer $\absr{f}_i$ satisfies a set of positive (negative) examples in $\pex$ ($\nex$):
\[ \textit{satI}^+(\absr{f}_{i}, \pex): \forall {\langle \langle a_1, \ldots, a_n \rangle, c' \rangle} \in \pex \, .~c' \in \gamma_{\pankaj{i}}(\absr{f}_{i}(a_1, \ldots, a_{n}))  \]
\[ \textit{satI}^-(\absr{f}_{i}, \nex): \forall {\langle \langle a_1, \ldots, a_{n} \rangle, c' \rangle} \in \nex \, .~c' \notin \gamma_{\pankaj{i}}(\absr{f}_{i}(a_1, \ldots, a_{n}))  \]

Next, we ``lift" these conditions to describe the predicate $sat^+$ to capture the condition that the reduced $\Lang$-transformer $\b{\dfuns}$ satisfies all examples in $\pex$:
\[ \textit{sat}^+(\langle \abstr{f}{1} \ldots \abstr{f}{n} \rangle, \pex): \bigwedge_{i = 1}^{n} \textit{satI}^+(\abstr{f}{i}, \pex) \]

and, that for each example in $\nex$, at least one component $\Lang$-transformer fails to satisfy the example: 
\[ \textit{sat}^-(\langle \abstr{f}{1} \ldots \abstr{f}{n} \rangle, \nex): \forall {\langle \langle a_1, \ldots, a_{n}\rangle, c' \rangle} \in \nex.\  \exists A_i\in \mathcal{D} .\ c' \not\in \gamma_{\pankaj{i}}(\abstr{f}{i}(a_1, \ldots, a_{n}))  \] 



We now use the above interfaces to construct checks for precision. Given a candidate $\Lang$-transformer $\b{\dfuns}$, the following check attempts to find a witness $\Lang$-transformer $\b{\absr{h}_1, \absr{h}_2, \dots, \absr{h}_n}$ and a negative counterexample $\b{\b{\args}, c'}$ such that:
\begin{itemize}
	\item the witness $\Lang$-transformer $\b{\absr{h}_1, \absr{h}_2, \dots, \absr{h}_n}$ includes all the positive examples in $\pex$;
	\item the witness $\Lang$-transformer excludes the negative example $\b{\b{\args}, c'}$ as well as the current set of negative examples $\nex$;
	\item the current $\Lang$-transformer $\b{\dfuns}$ does not exclude the negative example $\b{\b{\args}, c'}$.
\end{itemize}

This property can be formalized as follows:


\begin{align}
\label{eq:ReducedCheckPrecisionNaive}
&\exists  \langle \absr{h}_{1} \ldots \absr{h}_{n}\rangle, \langle \langle a_1, \ldots, a_{n}\rangle, c' \rangle\ \textrm{such that},\nonumber\\
&\text{\fcolorbox{red}{red!20}{$sat^+(\langle \absr{h}_{1}, \ldots, \absr{h}_{n} \rangle, \pex)$}}\ \land \nonumber\\
&\text{\fcolorbox{green!30!black}{green!20}{$sat^-(\langle \absr{h}_{1}, \ldots, \absr{h}_{n} \rangle, \nex \cup \{\langle\langle a_1, \ldots, a_{n}\rangle, c' \rangle \})$}}\ \land \nonumber \\
&\text{\fcolorbox{blue}{blue!20}{$\neg sat^-( \langle \absr{f}_{1}, \ldots, \absr{f}_{n} \rangle, \{\langle\langle a_1, \ldots, a_{n}\rangle, c' \rangle\})$}}
\end{align}



As discussed in \Cref{sec:product-overview}, the above check is not practical because it attempts to synthesize a set of $n$ functions $\b{\abs{h}_1, \dots, \abs{h}_n}$ in a single synthesis call. Instead, we define the 1-precision check that only attempts to synthesize one witness component transformer $\abs{h}_i$ at a time. Let us discuss how we can modify each term in the above precision check for 1-precision checking:

\begin{itemize}
	\item \text{\fcolorbox{red}{red!20}{$sat^+(\langle \absr{h}_{1}, \ldots, \absr{h}_{n}\rangle, \pex)$}} This term can be modified to the following for 1-precision check:
	\begin{equation}    
	sat^+( \langle \abstr{f}{1}, \ldots, \abstr{f}{(i-1)}, {\color{red}\abstr{h}{i}}, \abstr{f}{(i+1)}, \ldots, \abstr{f}{n} \rangle, \pex) \label{optPartA1}
	\end{equation}
	Here, we are only trying to synthesize one component transformer, i.e.,  {\color{red}$\abstr{h}{i}$}. Furthermore, from \Cref{optPartA1}, because all the component transformers except {\color{red}$\abstr{h}{i}$} already satisfy $\pex$ (by construction), we can remove
     $satI^+(\abstr{f}{1}, \pex)$, $\ldots$, $satI^+(\abstr{f}{(i-1))}, \pex)$, $satI^+(\abstr{f}{(i+1)}, \pex)$, $\ldots$, $satI^+(\abstr{f}{n}, \pex)$.
     
	This simplification yields the following equation:
	\begin{equation}
	satI^+(\abstr{h}{i}, \pex) \label{optPartA}
	\end{equation}
	
	\item \fcolorbox{green!30!black}{green!20}{$sat^-(\langle \absr{h}_{1}, \ldots, \absr{h}_{n} \rangle, \nex \cup \{\langle\langle a_1, \ldots, a_{n}\rangle, c' \rangle \})$}: This equation can be simplified to the following constraint.
	\begin{equation}
	\mathtt{sat}^-( {\langle \abstr{f}{1}, \ldots, \abstr{f}{(i-1)}, {\color{red}\abstr{h}{i}}, \abstr{f}{(i+1)}, \ldots, \abstr{f}{n}\rangle}, \nex \cup \{\langle\langle a_1, \ldots, a_{n}\rangle, c' \rangle \}) \label{optPartB1}
	\end{equation}
	
	Because all other domain transformers except $\absr{h}_i$ are held to their current definitions, the above equation can be written as the following simply by changing \nex to $\nex_i$,
	
	\begin{equation}
	\mathtt{satI}^-({\color{red}\abstr{h}{i}}, \nex_i \cup \{\langle\langle a_1, \ldots, a_{n}\rangle, c' \rangle \})
	\end{equation}
	
	\item \fcolorbox{blue}{blue!20}{$\neg sat^-( \langle \absr{f}_{1}, \ldots, \absr{f}_{n} \rangle, \{\langle\langle a_1, \ldots, a_{n}\rangle, c' \rangle\})$}: This equation does not undergo any changes because it does not involve the witness $\Lang$-transformer.
\end{itemize}

Finally, our 1-precision check for reduced product transformer can be formalized as:

\begin{align}
	\exists   {\color{red}\absr{h}_{i}}, \langle \langle a_1, \ldots, a_{n}\rangle, c' \rangle, \textrm{s.t.}\ & 
	  satI^+( {\color{red}\abstr{h}{i}}, \pex)\ \land  \nonumber  \\ 
	& satI^-({\color{red}\abstr{h}{i}}, \nex_i \cup \{\langle\langle a_1, \ldots, a_{n}\rangle, c' \rangle \})\ \land \label{eq:ReducedCheckPrecision}  \\ 
	& \neg sat^-( \langle \abstr{f}{1}, \ldots, \abstr{f}{n}\rangle, \{\langle\langle a_1, \ldots, a_{n}\rangle, c' \rangle\}) \nonumber
\end{align}

We will use the following interface function \textsc{CheckPrecision}, which implements the above 1-precision check:
\begin{align}
	\textsc{CheckPrecision}(&\langle \absr{f}_{1} \ldots \absr{f}_{n} \rangle, f, i, \pex, \nex_i) = \\
	&\begin{cases} \nonumber
		\textit{False}, \langle\langle a_1, \ldots, a_{n}\rangle, c' \rangle  & \textrm{if Eqn~\ref{eq:ReducedCheckPrecision} is SAT}\\
		\textit{True}, \_ & \text{otherwise}
	\end{cases}
\end{align}

\smallskip
\noindent
\fbox{\parbox{.98\linewidth}{
Please note that the precision check on the $i^{th}$ component transformer is conditioned on all the other component transformers $\{\absr{f}_1, \dots, \absr{f}_{i-1}, \absr{f}_{i+1}, \dots, \absr{f}_n\}$. The 1-precision status of $\absr{f}_i$ may change if any of the other component transformers change. 
}
}

\subsection{Synthesis}
\label{Se:Synthesis}

Given a set of positive examples $\pex$ and a set of negative examples $\nex_i$ for a component domain $A_i$, we attempt to synthesize a component transformer $\absr{f}_i$ that is consistent with these examples:

\[
  \exists \absr{f}_i \in \Lang_i \,.\, {\textit{satI}^+(\absr{f}_i, \pex)} \land {\textit{satI}^-(\absr{f}_i, \nex_i)}
\]

Such component transformers are combined into a reduced product transformer, $\absr{f}: \b{\dfuns}$. The reduced product transformer is expressed in a language $\Lang:\b{\Lang_1, \Lang_2, \dots, \Lang_n}$.

Due to the syntactic constraints of the DSL $\Lang$ that is used to express the abstract transformers, the synthesized reduced transformer is an overapproximation of the ideal reduced transformer, $\absr{\widehat{f}}$. 

\begin{figure}[t]
\centering
\includegraphics[scale=0.4]{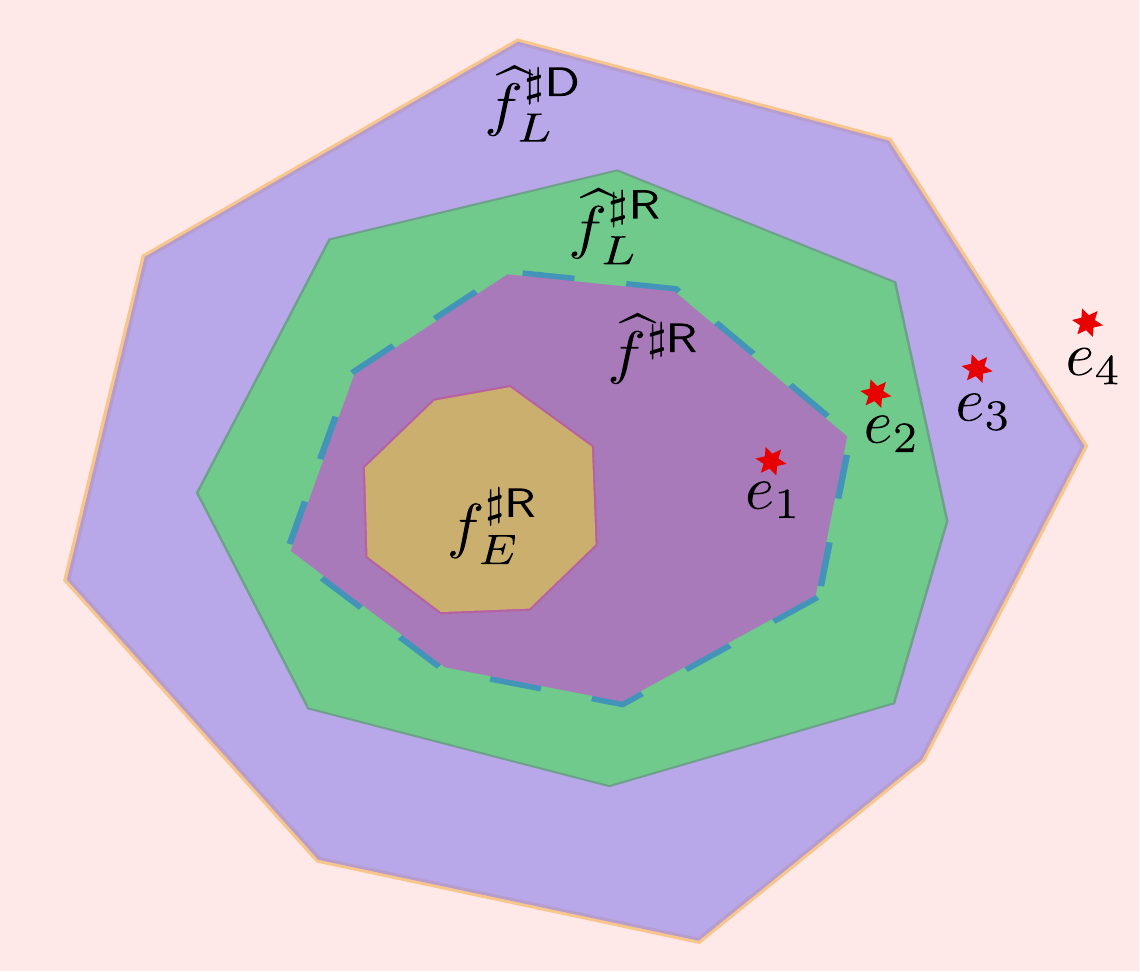}
\caption{Different configurations of negative examples
\label{fig:inconsistent_negex}}
\end{figure}

\Cref{fig:inconsistent_negex} shows different transformers, which classify the space of examples in different regions:
\begin{itemize}
  \item
    $\absr{f}_E$: This transformer is a candidate $\Lang$-transformer that satisfies the set of positive and negatives examples in the set of examples $E$.
  \item
    $\absr{\widehat{f}}$ is the ideal transformer for the given reduced-product domain (\Cref{eq:bestAbsTrans}).
    All examples within this region, like $e_1$, can be checked to be a positive example with a simple satisfiability query (discussed below); such examples will ``expand'' the synthesized transformer to make it sound.
  \item
    $\absr{\widehat{f}}_{\Lang}$ is a best reduced-product $\Lang$-transformer for the given reduced-product domain $A$;
    this transformer is one of the possible transformers that we would like to synthesize.
    Hence, examples like $e_2$ should be treated as positive examples for the complete product domain, and examples like $e_3$ should be treated as negative examples for this reduced-product transformer.
  \item
    $\absd{\widehat{f}}_{\Lang}$ is a best direct-product $\Lang$-transformer for the given direct-product domain $\mathcal{D}$ (see \Cref{sec:back:prod_dom}).
    All examples outside this region (pink zone), such as $e_4$, are clearly negative examples;
    such examples will ``shrink'' the synthesized transformer to make it more precise;
\end{itemize}

However, there does not exist any test to distinguish between examples $e_2$ and $e_3$.
Note that with the current state of the candidate transformer $\abs{f}_E$, both $e_2$ and $e_3$ can be emitted as negative examples (by \Cref{eq:ReducedCheckPrecision}). The example $e_2$ cannot be validated to be a positive example because the positive (counter)examples are generated with respect to \Cref{eq:ReducedCheckSoundness}, which essentially uses the definition the ideal transformer
$\absr{\widehat{f}}$.
At the same time, keeping $e_2$ as a negative example in $\nex$ would prevent us from synthesizing the desired transformer
$\absr{\widehat{f}}_{\Lang}$.

We resolve this problem as follows: if, at any stage, the synthesis problem turns unsatisfiable, we drop a \textit{minimal} set of negative examples that make synthesis feasible. The reason for dropping the smallest number of negative examples is inspired by Occam's razor. We achieve it by devising a strategy for \textsc{MaxSynth} in a synthesis domain---it is an analog of \textsc{MaxSAT} for the satisfiability domain.
\textsc{MaxSynth} solves a synthesis task by satisfying all hard constraints, while satisfying the maximum number of, but not necessarily all, soft constraints.
(If the hard constraints are unsatisfiable, \textsc{MaxSynth} returns $\bot$.)

We modify our synthesis task for the \textsc{MaxSynth} formulation where the satisfaction of positive examples are hard constraints and satisfaction of negative examples are treated as soft constraints, that is, we discount the smallest possible set of negative examples $\delta$ from $\nex$ such that synthesis becomes feasible.

\[ \exists \absr{f}_i \in \Lang_i \,.\, \overbrace{\textit{satI}^+(\absr{f}_i, \pex)}^{\textit{hard}} \land \overbrace{\textit{satI}^-(\absr{f}_i, \nex_i)}^{\textit{soft}} \]

We can also formulate it in terms of the negative examples that are dropped, $\delta$, as follows:

    \begin{align}    
      &\textsc{MaxSynthAll}(\pex, \nex_i) = \\ \nonumber
      & \begin{cases}
          \langle f^\sharp_{1}, \ldots, f^\sharp_{n} \rangle, \delta  &\textrm{if}~\exists \langle f^\sharp_{1}, \ldots, f^\sharp_{n} \rangle, \delta. \  sat^+(\langle f^\sharp_{1}, \ldots, f^\sharp_{n} \rangle, \pex)  \\
          &\land sat^-(\langle f^\sharp_{1}, \ldots, f^\sharp_{n} \rangle, \nex_i \setminus \delta), \\&\textit{where $\delta$ is minimal},\\
           \bot  &\mbox{otherwise}
       \end{cases}
    \end{align}

For 1-precision, the following definition of \textsc{MaxSynth} is sufficient, where the query is only over one component domain transformer.

    \begin{align}    
      &\textsc{MaxSynth}(\pex, \nex_i) =
       \begin{cases}
           \abst{f}{E}, \delta  &\textrm{if}~\exists \abst{f}{E}, \delta. \  satI^+(\abst{f}{E}, \pex)  \land \\
           &satI^-(\abst{f}{E}, \nex_i \setminus \delta), \\&\textit{where $\delta$ is minimal},\\
           \bot  &\mbox{otherwise}
       \end{cases}
    \end{align}

\paragraph{Optimizations.}
First, as discussed above, the examples obtained from the precision check are \textit{speculatively} treated as negative examples, but may be dropped via the \textsc{MaxSynth} query.
However, certain negative examples (like $e_1$ in \Cref{fig:inconsistent_negex}) can be ascertained to be a positive example via a simple satisfiability check.
The query given below provides a way to validate an example $\b{\b{\args}, c'}$ as a positive example
for a concrete operation $f$ and an abstract domain with concretization operations $\b{\gamma_1, \dots, \gamma_n}$:
\[
  \textsc{CheckPos}(\b{\b{a_1, \dots, a_n}, c'}) \equiv \exists c.\ c \in \gamma_1(a_1) \land \dots \land c \in \gamma_n(a_n) \land c' = {f}(c)
\]
If the above check succeeds, this example can be added to the global set of positive examples, $\pex$.

Similarly, any example that is not satisfied by the transformer for direct-product (like $e_4$ in \Cref{fig:inconsistent_negex}) is certainly a negative example.
This observation allows us to maintain a special set of negative examples, $E_i^{D-}$, that contain such \textit{surely} negative examples. The synthesis query can include such examples as hard constraints. 
\[ \exists \absr{f}_i \in \Lang_i \,.\, \overbrace{\textit{satI}^+(\absr{f}_i, \pex) \land \textit{satI}^-(\absr{f}_i, E_i^{D-})}^{\textit{hard}} \land \overbrace{\textit{satI}^-(\absr{f}_i, \nex_i \setminus E_i^{D-})}^{\textit{soft}} \]

To simplify matters, we do not show these optimizations in the statement of the core algorithm (\Cref{algo:synthRed}).


\subsection{Core Algorithm}

\Cref{algo:synthRed} shows our complete algorithm (sans optimizations).
It synthesizes a best (sound and maximally 1-precise in $\Lang$) reduced-product $\Lang$-transformer for a concrete function $f$ with respect to a product domain of $n$ component domains, $A_1 \times A_2 \times \dots A_n$, where the join, concretization, and abstraction operations are $\b{\sqcup_1,\sqcup_2,\ldots,\sqcup_n}, \b{\gamma_1,\gamma_2,\ldots,\gamma_n}$, and $\b{\alpha_1,\alpha_2,\ldots,\alpha_n}$, respectively.

\begin{algorithm2e}
	\caption {\textsc{SynthesizeReducedTransformer} \\ \hspace*{20pt}$(f, \D:A_1\times\ldots\times A_n, \b{\sqcup_1,\ldots,\sqcup_n}, \b{\gamma_1,\ldots,\gamma_n}, \b{\alpha_1,\ldots,\alpha_n}, \mathcal{G}, n)$\label{Alg:main} \label{algo:synthRed}}
	
	$\langle\abst{f}{1} ,\ldots ,\abst{f}{n}\rangle \gets ComputeDirectProduct()$\labelline{algoline:synthRed:initByDir}\\
	
	\ForEach{$k \in \{1,\ldots,n\}$}{
		$\textit{isSound}[k] \gets \textit{True}$ \labelline{algoline:synthRed:initBotS}\\
            $\textit{isPrecise}[k] \gets \textit{False}$ \labelline{algoline:synthRed:initBotP} \\
	}
	
	$\pex, \nex_1, \ldots, \nex_{n} \gets \func{InitializeExamples()}$ \labelline{algoline:synthRed:bootstrapping}
	
	$changed \gets \textit{True}$
	
	\While{$\textit{changed}$}{  \labelline{algoline:synthRed:outerwhile}
        $\textit{changed} \gets \textit{False}$ \\
		\ForEach{$k \in \{1,\ldots, n\}$}{ \labelline{algoline:synthRed:forLoop}
			\While{$\lnot \textit{isSound}[k] \lor \lnot \textit{isPrecise}[k]$} {\labelline{algoline:synthRed:innerWhile}

                        $\abst{f}{k} \gets \func{Synthesize}(\pex, \nex_k)$ \labelline{algoline:synthRed:synthesis}\\
                        \If{$\abst{f}{k} = \bot$}
					{
						$\abst{f}{k}, \delta \gets \func{MaxSynth}(\pex, \nex_k)$\labelline{algoline:synthRed:maxsatsynthesis}\\
						\eIf{$\abst{f}{k} \neq \bot $}{
							$\nex_k \gets \nex_k \setminus \delta$
						}
						{
							\Return $\textsc{Fail}$ \labelline{algoline:synthRed:synthesisFailure} 
						}
					}
			
				\If{$*$} { \labelline{algoline:synthRed:if-nondet} 
					$\textit{isSound}[k], e \gets \func{CheckSoundness}(f^\sharp_{k}, f)$  \labelline{algoline:synthRed:CallCheckSoundness} \\
					\If{$\lnot \textit{isSound}[k]$}{
						$\textit{isPrecise}[k] \gets \textit{False}$\\
						$\pex \gets \pex \cup \{e\}$ \labelline{algoline:synthRed:AdjustEPlus} \\
					}
				}
				\Else { \labelline{Alg:main:precise}
					$\textit{isPrecise}[k], e \gets \func{CheckPrecision}(\langle f^\sharp_{1}, \ldots, f^\sharp_{n} \rangle, k, \pex, \nex_k)$ \labelline{algoline:synthRed:CallCheckPrecision} \\
					\If{$\lnot \textit{isPrecise}[k]$}{
						{$\textit{isSound}[k] \gets \textit{False}$}\\
						$\nex_k \gets \nex_k \cup \{e\}$ \labelline{algoline:synthRed:AdjustEMinus}\\
                            \ForEach{$j \in \{1,\ldots, n\}$}{ \labelline{algoline:synthRed:forAllOthersSound}
                                $\textit{isPrecise}[j] \gets \textit{False}$\labelline{algoline:synthRed:forAllOthersSound1}
                            }
                        $\textit{changed} \gets \textit{True}$\\
					}
				}
			}\labelline{algoline:synthRed:innerWhileEnd}
		}\labelline{algoline:synthRed:forLoopEnd}
	}\labelline{algoline:synthRed:outerwhileEnd}
	\Return $\langle f^\sharp_{1}, \ldots, f^\sharp_{n} \rangle$  \labelline{algoline:synthRed:normalReturn}
\end{algorithm2e}

The algorithm maintains two classes of examples:
\begin{itemize}
    \item A \textit{global} set of positive examples, $\pex$;
    \item For each domain $A_i$, it maintains a \textit{private} set of negative examples, $\nex_i$.
\end{itemize}

Furthermore, for each domain $A_i$, it maintains two status flags, $\textit{isSound}[i]$ for soundness and $\textit{isPrecise}[i]$ for 1-precision; if any of these flags is false, it indicates the presence of new examples in the example sets that necessitate a call to the synthesis routine. The algorithm runs two CEGIS loops for the dual objectives of soundness and 1-precision. 

The algorithm primes the candidate reduced-product $\Lang$-transformer with a direct-product $\Lang$-transformer (\refline{algoline:synthRed:initByDir}).
This step ensures that all the $\Lang_i$-transformers are sound.
Each entry of the two arrays of flags $\textit{isSound}[.]$ and $\textit{isPrecise}[.]$ are initialized to \textit{true} and \textit{false}, respectively 
(\refline{algoline:synthRed:initBotS}, \refline{algoline:synthRed:initBotP}).
At \refline{algoline:synthRed:bootstrapping},
a set of positive examples and $n$ private sets of negative examples---one set for each component---are initialized to the sets of \textit{bootstrap} examples---optional examples that a user may provide to start the synthesis procedure. 

The while-loop from \refline{algoline:synthRed:outerwhile} to \refline{algoline:synthRed:outerwhileEnd} terminates only when a best reduced $\Lang$-transformer is found.
The subsequent \textsf{foreach}-loop (\refline{algoline:synthRed:forLoop}--\refline{algoline:synthRed:forLoopEnd}) iterates through every component domain $A_i$
to determine whether \textit{changing} the $\Lang_i$-component transformer for the domain yields a better reduced-product $\Lang$-transformer.
This for-loop finds a sound and 1-precise transformer for one domain before moving to the next domain. 

The inner while-loop from \refline{algoline:synthRed:innerWhile} to \refline{algoline:synthRed:innerWhileEnd} attempts to find a suitable
$\Lang_i$-transformer for a particular domain $A_i$:
at \refline{algoline:synthRed:if-nondet}, the algorithm makes a non-deterministic choice to invoke either a soundness check or a 1-precision check.
A positive example generated during the soundness check is added to the set \pex at \refline{algoline:synthRed:AdjustEPlus};
a negative example generated during the precision check is added to the negative-example set $\nex_k$ of the corresponding domain $k$ at \refline{algoline:synthRed:AdjustEMinus}.
Furthermore, the $\textit{isSound}[.]$ and $\textit{isPrecise}[.]$ flags for domain $k$ are set to false, indicating the necessity to synthesize a new transformer for domain $A_k$.
Interestingly, though $\pex$ is shared by all domains, any new positive example cannot invalidate the soundness status of the prior component transformers as all those transformers where already proven sound (the reason why the foreach-loop at \refline{algoline:synthRed:innerWhile} could break out of those domains). For a component transformer to be sound,  it must over-approximate a best transformer $\absr{\widehat{f}}_\Lang$ in the product domain modulo the language $\Lang$.
Because positive counterexamples can only be generated from $\absr{\widehat{f}}_\Lang$,
no new soundness examples can invalidate these component transformers.

There is an interesting case that needs to be handled when a new negative counterexample is discovered:
because a negative counterexample will force a new component transformer to be synthesized, and because the 1-precision of a component transformer $\absr{f}_i$ is conditioned on all other component transformers, the 1-precision status of all component transformers has to be invalidated when a new negative example is found
(\refline{algoline:synthRed:forAllOthersSound} to \refline{algoline:synthRed:forAllOthersSound1}).
Also, the status of the \textit{changed} flag is set to \textit{true},
to indicate that the algorithm is now required to cycle through all the domains again to re-synthesize all component $\Lang_i$-transformers with respect to the extended set of positive examples.

An $\Lang_k$-transformer for the domain $A_k$ is synthesized at \refline{algoline:synthRed:synthesis}, with respect to the global $\pex$ and local $\nex_i$ sets.
If synthesis fails, the algorithm calls \textsc{MaxSynth} (\refline{algoline:synthRed:maxsatsynthesis}) to drop a minimal number of negative examples, and produce a feasible $\Lang_k$-transformer.


\subsection{Theoretical Results}

\begin{lemma} The following invariants hold at \refline{algoline:synthRed:innerWhile} of \Cref{algo:synthRed}.
\label{lemma:invariant}
\begin{enumerate}
    \item All domain transformers $\absr{f}_i$, except for $\absr{f}_k$, are sound; 
    \item All domain transformers $\absr{f}_i$ are 1-precise if \texttt{isPrecise[i]} is \texttt{true}. 
\end{enumerate}
\begin{proof}
Invariant (1) holds due to the following reasons:
\begin{itemize}
    \item All transformers are sound to begin with (because the candidate reduced product is initialized with the direct product at \refline{algoline:synthRed:initByDir})
    \item The while-loop iteration (\refline{algoline:synthRed:innerWhile} to \refline{algoline:synthRed:innerWhileEnd}) corresponding to a component domain $k$ can terminate
    only if the respective transformer is sound and 1-precise;\footnote{
    The while-loop in \refline{algoline:synthRed:innerWhile} to \refline{algoline:synthRed:innerWhileEnd} can terminate on 
    \refline{algoline:synthRed:synthesisFailure} if synthesis fails.
    However, in that case control does not return to \refline{algoline:synthRed:innerWhile}.
    }
    because the soundness status of a transformer is not conditioned on others, it does not change due to synthesis of new transformers.
\end{itemize}
Invariant (2) holds due to the following reasons:
\begin{itemize}
    \item The flag \texttt{isPrecise[i]} is initialized to \texttt{False} to begin with (\refline{algoline:synthRed:initBotP});
    \item For the $k^{th}$ component transformer (the current loop iteration is at $k$), the 1-precision status can be invalidated due to the generation of new negative examples; \texttt{isPrecise[k]} is updated accordingly at \refline{algoline:synthRed:CallCheckPrecision};
    \item For all other component transformers, the 1-precision status of a component transformer is conditioned on the status of all other component transformers. \refline{algoline:synthRed:forAllOthersSound} to \refline{algoline:synthRed:forAllOthersSound1} invalidates the \texttt{isPrecise[.]} status of all component transformers whenever a negative counterexample is found (which will force a new transformer to be synthesized for each component).
\end{itemize}
\end{proof}
\qed
\end{lemma}

\begin{theorem}[Soundness]
Any reduced product transformer generated by \Cref{algo:synthRed} (at \refline{algoline:synthRed:normalReturn}) will be sound; that is, all the component transformers $\absr{f}_i$ are sound.

\begin{proof}
  This property holds because of invariant (1) of \Cref{lemma:invariant}, and because the $k^{\textit{th}}$ iteration of the foreach-loop from line \refline{algoline:synthRed:forLoop} to   \refline{algoline:synthRed:forLoopEnd}  
  can progress to the next iteration only when the transformer $\absr{f}_k$ is sound (cf.\ \refline{algoline:synthRed:innerWhile}).
\end{proof}
\qed
\end{theorem}

\begin{theorem}[Precision]
Any reduced product transformer generated by \Cref{algo:synthRed} (at \refline{algoline:synthRed:normalReturn}) will be 1-precise, that is, each of the component transformers $\absr{f}_i$ is 1-precise.

\begin{proof}
  This property holds because the while-loop from \refline{algoline:synthRed:outerwhile} to \refline{algoline:synthRed:outerwhileEnd} can exit at \refline{algoline:synthRed:outerwhile} only if all component transformers are proved to be 1-precise.
  This property can be established via invariant (2) of \Cref{lemma:invariant}:
  \begin{itemize}
      \item the $k^{th}$ iteration of the foreach-loop from line \refline{algoline:synthRed:forLoop} to   \refline{algoline:synthRed:forLoopEnd} can progress to the next iteration only when $\absr{f}_k$ is 1-precise;
      \item for component transformers other than $k$, whenever their \texttt{isPrecise[k]} flag is set to \texttt{false}, the \texttt{changed} flag is set to \texttt{true}, which forces the 1-precision check at \refline{algoline:synthRed:CallCheckPrecision} to be revisited for every component transformer.
  \end{itemize}
\end{proof}
\qed
\end{theorem}

The innermost loop (\refline{algoline:synthRed:innerWhile}--\refline{algoline:synthRed:innerWhileEnd}) in \Cref{algo:synthRed} is guaranteed to terminate if the component-domain DSLs $\Lang_1, \Lang_2, \ldots, \Lang_n$ are all finite, and we use a fair scheduler (such as a round-robin scheduler) to resolve the non-deterministic choice at \refline{algoline:synthRed:if-nondet}.
The proof of termination is similar to the one given by Kalita et al.~\cite[Thm.\ 4.4]{amurth}.

In cases where not all of the component-domain DSLs are finite, the
\textsf{for} loop (\refline{algoline:synthRed:forLoop}--\refline{algoline:synthRed:forLoopEnd}) runs for $\mathsf{n}$ times,
but the outermost loop (\refline{algoline:synthRed:outerwhile}--\refline{algoline:synthRed:outerwhileEnd}) may not terminate. However,
in our experiments, we did not encounter any instances of non-termination.

\section{Case Studies}
\label{sec:case_studies}

\tool is implemented in Python, and uses the Sketch engine~\cite{sketch} (v.\ 1.7.5) for the synthesis tasks.
The experiments were conducted on an Intel(R) Core(TM) i7-8700 CPU @ 3.20GHz CPU with 32GB RAM, running Ubuntu 18.04. To finitize our language, we unroll the recursive productions in our DSL to at most an unrolling depth of three.
\tool was given a timeout of 600 seconds for each call to Sketch.
All timing results presented in this section report the median of three runs. 



For our case-studies, we considered three reduced-product domains: an integer domain for even-intervals and odd-intervals (described in \Cref{sec:back:prod_dom}), along with two popular string product domains, SAFE~\cite{FOOL:LWJCR12}, and JSAI~\cite{DBLP:conf/sigsoft/KashyapDKWGSWH14}, which are available in the \safestr JavaScript analysis engine~\cite{TACAS:SAFEstr17}. Interestingly, perhaps due to the difficulty of establishing the soundness of reduced-product transformers,
\safestr uses the reduced-product transformer only for \concat. For all other operations, such as \tolower, \toupper, \trim, \contains, and \charat, it relies on the direct-product transformers.
Using appropriate DSLs, \tool could infer \textit{more precise reduced-product transformers} for many of these operations.


\subsection{Case Study I: Reduced $\Lang$-Transformers for the Reduced Product of the Odd-Interval and Even-Interval Domains}


We used \tool to implement reduced $\Lang$-transformers for four operations: 
increment, addition, subtraction, and absolute value.
The DSL used is provided in \Cref{eq:intervalDSL}. The transformers synthesized by \tool are shown below:

\begin{align}
     \rept{add_{O\times E}}{\mathtt{o_1, e_1}} {\mathtt{o_2, e_2}} =&\nonumber \\ 
    &\hspace{-3.5cm} \langle[\colorbox{red!20}{$\mathtt{max(o_2.l + e_1.l,\ o_1.l + e_2.l)}$, $\mathtt{max(o_1.r + o_2.r,\ e_1.r + e_2.r) - 1}$}],\nonumber\\
    &\hspace{-3.3cm} [\colorbox{blue!20}{$\mathtt{max(o_1.l + o_1.l,\ e_1.l + e_2.l)}$, $\mathtt{max(o_1.r + e_2.r,\ o_2.r + e_1.r) - 1}$}]\rangle \label{eq:addReduced}
\end{align}

\begin{align}
    \hspace{-1.1cm}\rept{sub_{O\times E}}{o_1, e_1}{o_2, e_2} =& \nonumber\\
    &\hspace{-3cm} \langle[\colorbox{red!20}{$\mathtt{max(o_1.l-e_2.r, e_1.l-o_2.r)}$, $\mathtt{min(e_1.r-o_2.l, o_1.r-e_2.l)}$}], \nonumber \\ 
    &\hspace{-2.8cm}[\colorbox{blue!20}{$\mathtt{max(o_1.l - o_2.r, e_1.l-e_2.r)}$, $\mathtt{min(o_1.r-o_2.l, e_1.r-e_2.l})$}]\rangle \label{eq:subReduced}
\end{align}

\begin{align}
    \mathtt{inc^{\sharp R}_{O\times E}(\langle o, e\rangle)} &= \langle\mathtt{[\colorbox{red!20}{e.l + 1, e.r+1}]}, \mathtt{[\colorbox{blue!20}{o.l + 1, o.r+1}]}\rangle \label{eq:incRedOddExperiment}
\end{align}

\begin{align}
    \repo{abs_{O\times E}}{o, e} = \langle& [\colorbox{red!20}{$\mathtt{max(max(-1, a.l), -a.r)}$, $\mathtt{max(-a.l, a.r)}$}],\nonumber \\ &[\colorbox{blue!20}{$\mathtt{max(max(0, a.l), -a.r)}$, $\mathtt{max(-a.l, a.r)}$}] \rangle \label{eq:absReduced}
\end{align}

\noindent
For these operations, \tool took the following times to synthesize the reduced-product $\Lang_{O\times E}$-transformers: 1871s for \texttt{add}, 2466s for \texttt{sub}, 2109s for \texttt{inc} and 2312s for \texttt{abs}. 

\begin{figure}[t]
	\centering
	\begin{subfigure}{.4\linewidth}
		\centering
		\includegraphics[scale=1]{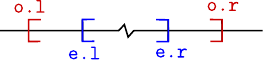}
		\caption{\label{fig:inc1}}
	\end{subfigure}
\hfill
	\begin{subfigure}{.4\linewidth}
		\centering
		\includegraphics[scale=1]{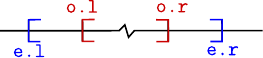}
		\caption{\label{fig:inc2}}
		\vspace{5mm}
	\end{subfigure}
	\begin{subfigure}{.4\linewidth}
		\centering
		\includegraphics[scale=1]{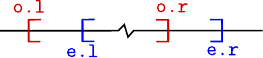}
		\caption{\label{fig:inc3}}
	\end{subfigure}
\hfill
	\begin{subfigure}{.4\linewidth}
		\centering
		\includegraphics[scale=1]{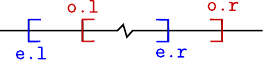}
		\caption{\label{fig:inc4}}
	\end{subfigure}
	\caption{Different configurations for limits of  odd and even intervals\label{fig:incrementConfig}}
\end{figure}

As one can see, the transformers tend to get complex even for simple concrete operations.
For example, consider how the left-limit for the odd-interval domain is computed in the reduced $\Lang_{O\times E}$-transformer for subtraction (see \Cref{eq:subReduced}).
Note that the components of the odd intervals and even intervals can appear in any of the possible configurations shown in \Cref{fig:incrementConfig}.
Subtracting the even-interval domain's right-limit from the odd-interval domain's left-limit (and vice versa) is sound, and also produces a value that is odd.
Moreover, taking the maximum preserves soundness because it makes a choice between two limits that are both sound, and also selects the higher of the two left limits, thereby choosing the more-precise option.
Hence, by cleverly choosing between two carefully constructed sound left limits based on the information from \textit{both} the odd-interval and even-interval domains, \tool is able to construct the most precise reduced $\Lang_{O\times E}$-transformer.
In contrast, the direct-product $\Lang_{O\times E}$-transformer for this operation is less precise:
\begin{align}
\dipt{sub_{O \times E}}{o_1, e_1}{o_2, e_2} = \langle &[\colorbox{red!20}{$\mathtt{o_1.l - o_2.r - 1}$,\ $\mathtt{o_1.r - o_2.l + 1}$}],\nonumber\\   &[\colorbox{blue!20}{$\mathtt{e_1.l - e_2.r}$,\ $\mathtt{e_1.r - e_2.l}$}] \rangle\label{eq:subDirect}
\end{align}

%
%
%
%
%
%

%


\subsection{Case Study II: Reduced $\Lang$-Transformers for the \safe Domain}

\subsubsection{Domain Description.}
\safe is a reduced product of two string domains, \ssk and \no. 

\paragraph{String Set Domain (\ssk).}
This string domain precisely represents a set of bounded ($k \geq 1$) concrete strings~\cite{TACAS:SAFEstr17}. It is parametric on $k$, that is, the size of the string set. The abstraction ($\alpha$) and concretization ($\gamma$) functions of this domain are as follows:
\begin{align}
	\alpha_{\ssk}(C) &= 	 
		\begin{cases}
			C & |C| \leq k \\
			\top_{\ssk} & otherwise
		\end{cases}\\[1em]
	\gamma_{\ssk}(A) &= 
		\begin{cases}
			A & A \neq \top_{\ssk} \\
			\Sigma^* & otherwise
		\end{cases}
\end{align}

\paragraph{Number-or-Other (\no) Domain.}
This domain is another string domain that is used in \cite{TACAS:SAFEstr17}.
It keeps track of a few weak properties of strings, i.e., whether
the string is a \textit{numeric string} or some \textit{other string}. 
Numbers, e.g., $-3, 0, 53$, along with \texttt{NaN} are treated as \textit{numeric strings} (\textsc{NumStr}), and the rest are considered to be \textit{other strings} (\textsc{OtherStr}).

\begin{figure}[t]\centering
    \begin{subfigure}[b]{.4\linewidth}\centering
        \includegraphics[scale=.4]{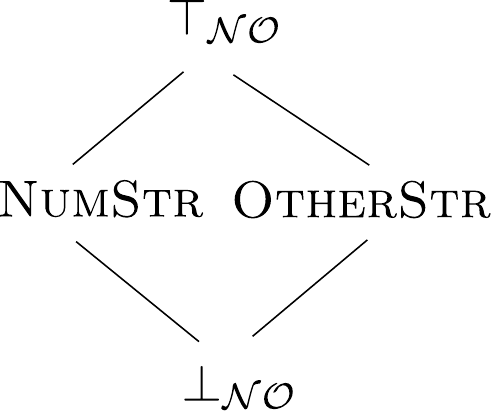}
        \caption{Lattice for \no\label{fig:noLattice}}
    \end{subfigure}
    \begin{subfigure}[b]{.5\linewidth}\centering
        \includegraphics[scale=.42]{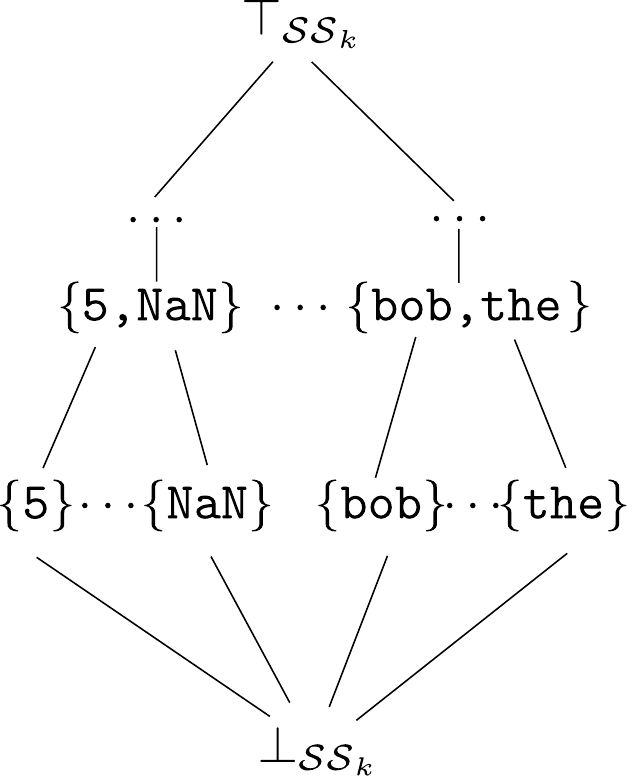}
        \caption{Lattice for \ssk\label{fig:sskLattice}}
    \end{subfigure}
    \caption{Lattices for \no and \ssk domains}
\end{figure}

The \safe domain is a reduced product of the \no and \ssk domains. 
\Cref{fig:noLattice}, \Cref{fig:sskLattice} show the lattice structures for the \ssk, and \no domains, respectively. 

\subsubsection{DSL used.}
The DSL $\Lang_\safe$ is essentially the same as the DSL used by Kalita et al.~ \cite{amurth,amurthArtifact}.

\subsubsection{The \concat operation.}
\Cref{fig:concatSAFE} shows the pseudocode for the $\Lang_\safe$-transformer for the \concat operation synthesized by \tool.
The synthesized version of the $\Lang_\safe$-transformer is semantically equivalent to the manually written version available in \safestr. 

The arguments, $arg_1$ and $arg_2$, are abstract values in the \safe domain, where the \texttt{ssk} and \texttt{no} fields of each abstract value represent the abstract values in the \ssk and \no domains.
The $\Lang_\safe$-transformer operates as follows: if the \texttt{ssk} component of both arguments are not $\top_{\ssk}$ or $\bot_{\ssk}$ in \ssk, the $\Lang_\safe$-transformer iterates over every string and concatenates the respective strings.
If the cardinality of the resultant set ($strset$) exceeds the maximum set cardinality for the \ssk domain ($k$), then the \texttt{ssk} component of the return value will be $\top_{\ssk}$.
The resultant \texttt{sset} can be used to create a precise abstract value for \no domain. In case the \texttt{ssk} component is $\top_{\ssk}$ or $\bot_{\ssk}$, the $\Lang_\safe$-transformer invokes the respective domain transformers for both component domains (\refline{code:concatDirectSafe}).

\begin{figure}[t]
\centering
\begin{minipage}{.9\linewidth}
\begin{minted}[autogobble, fontsize=\small, tabsize=2, escapeinside=||, breaklines, linenos]{java}
|concat$_{\safe}^\sharp$|(|$\mathtt{arg_1, arg_2}$|) {
	if(|$\mathtt{arg_1.ssk} \not\in \{\top_{\ssk},\bot_{\ssk}\}  \land \mathtt{arg_2.ssk} \not\in \{\top_{\ssk},\bot_{\ssk}\} $|)
	{
		|$\mathtt{sset \gets \emptyset}$|
		for(|$\mathtt{x \gets arg_1.ssk}$|)
			for(|$\mathtt{y \gets arg_2.ssk}$|)
				|$\mathtt{sset \gets sset \cup \{\concat(x, y)\}}$|
		|$\mathtt{out.no \gets \alpha_{\no}(sset)}$|
		|$\mathtt{out.ssk \gets (\mid sset\mid > k)\ ?\ \top_{\ssk}\ :\ \alpha_{\ssk}(sset)}$|
		return out
	} else {
		return |$\langle\mathtt{\concat_{\ssk}^\sharp(arg_1.ssk, arg_2.ssk), \concat_{\no}^\sharp(arg_1.no, arg_2.no)}\rangle$\labelline{code:concatDirectSafe}|
  }
}
\end{minted}
\end{minipage}
\caption{Reduced $\Lang_\safe$-transformer for the \concat operation \label{fig:concatSAFE}}
\end{figure}

\subsubsection{The \trim operation.}

\begin{figure}
    \includegraphics[scale=.3]{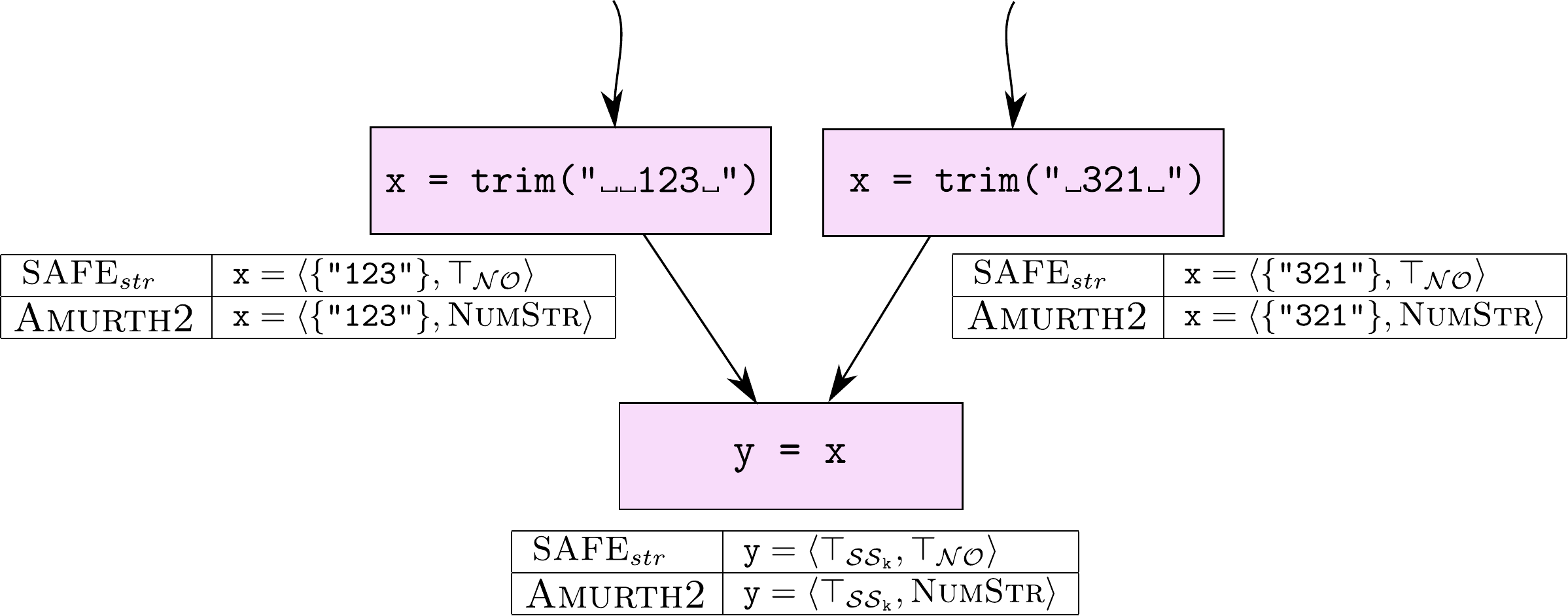}
    \caption{Showing differences in abstract values while using transformers from \safestr and \tool \label{fig:trimSafeDirVSRed}}
    \label{fig:enter-label}
\end{figure}

The concrete \trim operation removes leading and trailing whitespace characters from the provided string. For example, \texttt{trim("\textvisiblespace\textvisiblespace\textvisiblespace New\textvisiblespace York\textvisiblespace\textvisiblespace")} will result in \texttt{"New\textvisiblespace York"},
where `\textvisiblespace' represents a space character. 


\Cref{fig:trimSAFEDirect} shows the $\Lang_\safe$-transformer available in \safestr.
Consider an abstract value that has the singleton set $\{\texttt{"\textvisiblespace123\textvisiblespace\textvisiblespace"}\}$ as the \ssk component and \textsc{OtherStr} in the \no component.
In this case, the $\Lang_\safe$-transformer synthesized by \tool returns $\b{\{\texttt{"123"}\}, \textsc{NumStr}}$ for the \safe domain while the $\Lang_\safe$-transformer in \safestr returns $\b{\{\texttt{"123"}\}, \top_{\no}}$. Because the analysis will fetch the \textit{smaller} (meet) of the component abstract values, both the direct and reduced transformer still return the maximally precise solution. However, the program shown in \Cref{fig:trimSafeDirVSRed}
illustrates a case where the \trim operation appears on two different paths that meet at some program point. Assuming $k=1$ for the \ssk component domain, the analysis will lose precision if the direct-product is used (as is the case in \safestr):
in the provided example, the direct-product produces $\top_{\safe}$ while the reduced-product (as synthesized by \tool) infers it as a set of number strings (\textsc{NumStr}).


\begin{figure}[t]
\centering

	\begin{subfigure}[b]{.45\linewidth}
		\centering
 
		\begin{minted}[autogobble, fontsize=\small, tabsize=2, escapeinside=||, breaklines, linenos]{C}
		|$\mathtt{\absr{trim}_{\safe}}(\mathtt{arg_1})$| {
			if(|$\mathtt{arg_1.ssk} \not\in \{\top_{\ssk},\bot_{\ssk}\}$|) {
				|$\mathtt{sset \gets \emptyset}\labelline{code:trimSafe1}$|
				for(|$\mathtt{x \gets arg_1.ssk}$|)
					|$\mathtt{sset \gets sset \cup \{trim(x)\}}\labelline{code:trimSafeAp}$|
				|$\mathtt{out.no \gets \alpha_{\no}(sset)}$|
				|$\mathtt{out.ssk \gets \alpha_{\ssk}(sset)}$|
				return out |$\labelline{code:trimSafe2}$|
			} else {
				if(|$\mathtt{arg_1.no = \textsc{OtherStr}}\labelline{code:trimSafe3}$|)
					return |$\mathtt{\langle arg_1.ssk, \top_{\no} \rangle}$|
				else
					return |$\mathtt{arg_1}\labelline{code:trimSafe4}$|
			}
		}
		\end{minted}
		\caption{
  $\Lang_\safe$-transformer for \trim synthesized by \tool
             \label{fig:trimSAFE}}
	\end{subfigure}
	\hfill
	\begin{subfigure}[b]{.45\linewidth}
		\centering
		\begin{minted}[autogobble, fontsize=\small, tabsize=2, escapeinside=||, breaklines, linenos]{C}
|$\mathtt{\absd{trim}_{\safe}}(\mathtt{arg_1})$| {
  out |$\gets \mathtt{arg_1}$|
  if(|$\mathtt{arg_1.ssk} \not\in \{\top_{\ssk},\bot_{\ssk}\} $|) {
    |$\mathtt{sset \gets \emptyset}$| 
    for(|$\mathtt{x \gets arg_1.ssk}$|)
      |$\mathtt{sset \gets sset \cup \{trim(x)\}}$|
    |$\mathtt{out.ssk \gets \alpha_{\ssk}(sset)}$|
  }
  if(|$\mathtt{arg_1.no = \textsc{OtherStr}}\labelline{code:trimSafe3}$|)
    return |$\mathtt{\langle out.ssk, \top_{\no} \rangle}$|
  else
    return |$\mathtt{out}\labelline{code:trimSafe4}$|		
}
		\end{minted}
		\caption{
  Manually written $\Lang_\safe$-transformer for \trim found in \safestr
          \label{fig:trimSAFEDirect}}
	\end{subfigure}
    \caption{
    $\Lang_\safe$-transformers for \trim
    }
\end{figure}

\subsubsection{The \tolower operation.}

The concrete \tolower operation accepts a string and makes each character lowercase, e.g., \texttt{\tolower("Hello") = "hello"}.
However, any numeric string is left unchanged, except \texttt{NaN}.


\Cref{fig:toLowerSAFEDirect} shows the $\Lang_\safe$-transformer available in \safestr, which essentially performs a direct-product.
The following scenario describes a case where the reduced $\Lang_\safe$-transformer synthesized by \tool is more precise than the $\Lang_\safe$-transformer that \safestr implements.
Consider an abstract value that has the singleton set $\{\texttt{"123"}\}$ as the \ssk component.
On \tolower, the $\Lang_\safe$-transformer from \safestr (\Cref{fig:toLowerSAFEDirect}) returns $\top_{\no}$.
The $\Lang_\safe$-transformer synthesized by \tool
uses the code in
\refline{code:toLowSafe1} to \refline{code:toLowSafe2} in \Cref{fig:toLowerSAFE} to return \textsc{NumStr} for the \no domain, which is more precise.
This can affect the precision of the analysis for a reason similar to the case of \trim.  

\begin{figure}[t]

	\begin{subfigure}[b]{.45\linewidth}
		\centering
		\begin{minted}[autogobble, fontsize=\small, tabsize=2, escapeinside=||, breaklines, linenos]{C}
			|$\mathtt{\absr{toLower}_{\safe}(arg_1)}$| {
				if(|$\mathtt{arg_1.ssk} \not\in \{\top_{\ssk},\bot_{\ssk}\} $|) {
					|$\mathtt{sset \gets \emptyset}\labelline{code:toLowSafe1}$|
					for(|$\mathtt{x \gets arg_1.ssk}$|)
							|$\mathtt{sset \gets sset \cup \{toLower(x)\}}$|
					|$\mathtt{out.no \gets \alpha_{\no}(sset)}$|
					|$\mathtt{out.ssk \gets \alpha_{\ssk}(sset)}$|
					return out|$\labelline{code:toLowSafe2}$|
				} else {
					if(|$\mathtt{arg_1.no = \textsc{NumStr}} \labelline{code:toLowSafe3} \labelline{code:toLowSafeNum}$|)
						return |$\langle\mathtt{arg_1.ssk, \top_{\no}}\rangle$|
					else
						return |$\mathtt{arg_1}\labelline{code:toLowSafe4}$|
				}
			}
		\end{minted}
		\caption{
   $\Lang_\safe$-transformer for \tolower synthesized by \tool
                \label{fig:toLowerSAFE}
            }
	\end{subfigure}
	\hfill
	\begin{subfigure}[b]{.50\linewidth}
		\centering
    \begin{minted}[autogobble, fontsize=\small, tabsize=2, escapeinside=||, breaklines, linenos]{C}
        |$\mathtt{\absd{toLower}_{\safe}}(\mathtt{arg_1})$| {
            out |$\gets \mathtt{arg_1}$|
            if(|$\mathtt{arg_1.ssk} \not\in \{\top_{\ssk},\bot_{\ssk}\} $|) {
                |$\mathtt{sset \gets \emptyset}$| 
                for(|$\mathtt{x \gets arg_1.ssk}$|)
                    |$\mathtt{sset \gets sset \cup \{toLower(x)\}}$|
                |$\mathtt{out.ssk \gets \alpha_{\ssk}(sset)}$|
            }
            if(|$\mathtt{arg_1.no = \textsc{NumStr}} $|)
                    return |$\langle\mathtt{out.ssk, \top_{\no}}\rangle$|
                else
                    return out
        }
    \end{minted}
		\caption{
  Manually written $\Lang_\safe$-transformer for \tolower found in \safestr
  \label{fig:toLowerSAFEDirect}}
	\end{subfigure}
    \caption{$\Lang_\safe$-transformers for \tolower}
\end{figure}

\subsubsection{The \toupper operation.}
The concrete \toupper operation converts each lowercase character to its uppercase character.
The synthesized $\Lang_\safe$-transformer for \toupper (\Cref{fig:toUpperSafe}) is similar to the synthesized $\Lang_\safe$-transformer for \tolower. 
The $\Lang_\safe$-transformer available in \safestr is provided in \Cref{fig:toUpperSAFEDirect}.
Again, the $\Lang_\safe$-transformer synthesized by \tool is more precise than \Cref{fig:toUpperSAFEDirect}.
For example, on the concrete string \texttt{"NaN"}, \Cref{fig:toUpperSAFEDirect} will return $\top_{\no}$;
however, the reduced-product $\Lang_\safe$-transformer (\Cref{fig:toUpperSafe}) will return \textsc{OtherStr}, which is more precise.
This can affect the precision of the analysis for a reason similar to the case of \trim.  

\begin{figure}[t]\centering
	\begin{subfigure}[b]{.49\linewidth}
		\centering
		\begin{minted}[autogobble, fontsize=\small, tabsize=2, escapeinside=||, breaklines, linenos]{C}
		|$\mathtt{\absr{toUpper}_{\safe}}(\mathtt{arg_1})$| {
			if(|$\mathtt{arg_1.ssk} \not\in \{\top_{\ssk},\bot_{\ssk}\}$|) {
				|$\mathtt{sset \gets \emptyset}$|
				for(|$\mathtt{x \gets arg_1.ssk}$|)
					|$\mathtt{sset \gets sset \cup \{toUpper(x)\}}$|
				|$\mathtt{out.no \gets \alpha_{\no}(sset)}$|
				|$\mathtt{out.ssk \gets \alpha_{\ssk}(sset)}$|
				return out
			} else {
				if(|$\mathtt{arg_1.no = \textsc{NumStr}}$|)
					return |$\mathtt{\langle arg1.ssk, \top_{\no} \rangle}$|
				else
					return |$\mathtt{arg_1}$|
			}
		}
		\end{minted}
		\caption{
   $\Lang_\safe$-transformer for \toupper synthesized by \tool
           \label{fig:toUpperSafe}}
	\end{subfigure}
	\begin{subfigure}[b]{.49\linewidth}
		\centering
		\begin{minted}[autogobble, fontsize=\small, tabsize=2, escapeinside=||, breaklines, linenos]{C}
        |$\mathtt{\absd{toUpper}_{\safe}}(\mathtt{arg_1})$| {
          out |$\gets \mathtt{arg_1}$|
          if(|$\mathtt{arg_1.ssk} \not\in \{\top_{\ssk},\bot_{\ssk}\}$|) {
            |$\mathtt{sset \gets \emptyset}$| 
            for(|$\mathtt{x \gets arg_1.ssk}$|)
              |$\mathtt{sset \gets sset \cup \{toUpper(x)\}}$|
            |$\mathtt{out.ssk \gets \alpha_{\ssk}(sset)}$|
          }
          if(|$\mathtt{arg_1.no = \textsc{NumStr}} $|)
            return |$\langle\mathtt{out.ssk, \top_{\no}}\rangle$|
          else
            return out
        }
		\end{minted}
		\caption{
  Manually written $\Lang_\safe$-transformer for \toupper found in \safestr
            \label{fig:toUpperSAFEDirect}}
	\end{subfigure}
    \caption{$\Lang_\safe$-transformers for \toupper}
\end{figure}

\subsubsection{The \contains operation.}
The concrete \contains operation returns \texttt{true} if the string provided as the second argument is a contiguous substring of the first argument; otherwise, it returns \texttt{false}. 

In case of \contains, the reduced-product $\Lang_{\safe}$-transformer (synthesized by \tool) offers the same precision as the direct product transformer (available in \safestr). \Cref{fig:containsSAFE} shows the transformer synthesized by \tool for \contains in the \safe domain.

\begin{figure}[t]
		\centering
		\begin{minted}[autogobble, fontsize=\small, tabsize=1, escapeinside=||, breaklines, linenos]{C}
|$\mathtt{\absr{\contains}_{\safe}}(\mathtt{arg_1, arg_2})$| {
  if(|$\mathtt{arg_1.ssk \not\in \{\top_{\ssk},\bot_{\ssk}\}} \land$  $\mathtt{arg_2.ssk \not\in \{\top_{\ssk},\bot_{\ssk}\}}$\labelline{code:containsSafe1}|) {
    fa = true; ex = false
    for |$x \gets \mathtt{arg_1.ssk}$|
      for |$y \gets \mathtt{arg_2.ssk}$|
        r |$\gets$| contains(x,y)
        fa |$\gets$| fa |$\land$| r
        ex |$\gets$| ex |$\lor$| r
    if(fa) return BoolTrue
    if(ex) return BoolTop
    return BoolFalse |\labelline{code:containsSafe2}|
  } else {
    return |$\mathtt{contains_{\ssk}^\sharp(arg_1.ssk, arg_2.ssk)} \sqcap \mathtt{contains_{\no}^\sharp(arg_1.no, arg_2.no)}$ \labelline{code:safeDirectContains}|  
  }
}
\end{minted}
\caption{$\Lang_\safe$-transformer for \contains synthesized by \tool \label{fig:containsSAFE}}
\end{figure}

\begin{figure}[t]
		\centering
		\begin{minted}[autogobble, fontsize=\footnotesize, tabsize=2, escapeinside=||, breaklines, linenos]{C}
		|$\mathtt{\absr{\charat}_{\safe}}(\mathtt{arg_1:SAFE, pos:NUM})$| {
			if(|$\mathtt{arg_1.ssk} \not\in \{\top_{\ssk}, \bot_{\ssk}\} \land \mathtt{pos} \not\in \{\top_{num}, \bot_{num}\}$|) {
				|$\mathtt{sset \gets \emptyset}\labelline{code:charatSafe1}$|
				for(|$\mathtt{x \gets arg_1.ssk}$|)
					|$\mathtt{sset \gets sset}\ \cup$ ($\mathtt{x.len \geq pos}$ ? $\mathtt{\{charAt(x, pos)\}: EMPTY)} \labelline{code:charatSafeAp}$|
				|$\mathtt{out.no \gets \alpha_{\no}(sset)}   \labelline{code:charatImprovedNOValue}$|
				|$\mathtt{out.ssk \gets \alpha_{\ssk}(sset)}$|
				return out |$\labelline{code:charatSafe2}$|
			} else {
				return |$\langle \mathtt{\abs{charAt}_{\ssk}(arg_1.ssk, pos)},$ $\mathtt{\abs{charAt}_{\no}(arg_1.no, pos)}\rangle$\labelline{code:safeDirectCharAt}|
			}
		}
		\end{minted}
		\caption{
   $\Lang_\safe$-transformer for \charat synthesized by \tool. The first argument $\mathtt{arg_1}$ is a value in the \safe domain, while $\mathtt{arg_2}$ is a value in a numeric abstract domain. $\mathtt{EMPTY}$ refers to an empty string. \label{fig:charatSAFE}}
\end{figure}

\subsubsection{The \charat operation.}
The concrete operation for \charat accepts two arguments, a string, and an index: it returns the character from the input string at the provided index. 
\Cref{fig:charatSAFE} shows the reduced $\Lang_\safe$-transformer synthesized by \tool.

Again, the $\Lang_\safe$-transformer synthesized by \tool is more precise than the $\Lang_\safe$-transformer provided by \safestr.
Consider what happens when the first argument is an abstract value that has the singleton set $\{\texttt{"ab12cd"}\}$ as the \ssk component and \textsc{OtherStr} in the \no component, and the second argument is the value $\{\texttt{3}\}$ in a numeric abstract domain.
Due to the limitations of the \no domain, it is impossible to return a precise answer (using \no alone).
However, one can obtain a more precise answer for the \no component when the string set from the \ssk domain is available, as evidenced by the code in \refline{code:charatSafe1} to \refline{code:charatSafe2} in  \Cref{fig:charatSAFE}.
(The assignment in \refline{code:charatImprovedNOValue} sets the \no component of the return value.)

\subsection{Case Study III: Reduced Transformers for the JSAI Domain}

\subsubsection{The JSAI Domain.}

\begin{figure}[t]\centering
    \begin{subfigure}[b]{.4\linewidth}\centering
        \includegraphics[scale=.4]{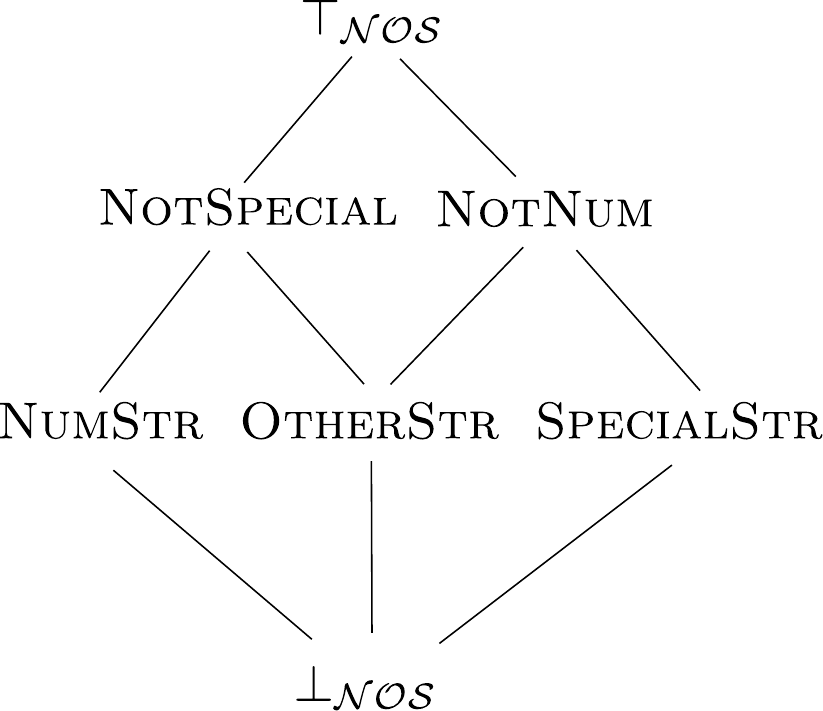}
        \caption{Lattice for \nos\label{fig:nosLattice}}
    \end{subfigure}
    \hfill
    \begin{subfigure}[b]{.4\linewidth}\centering
        \includegraphics[scale=.4]{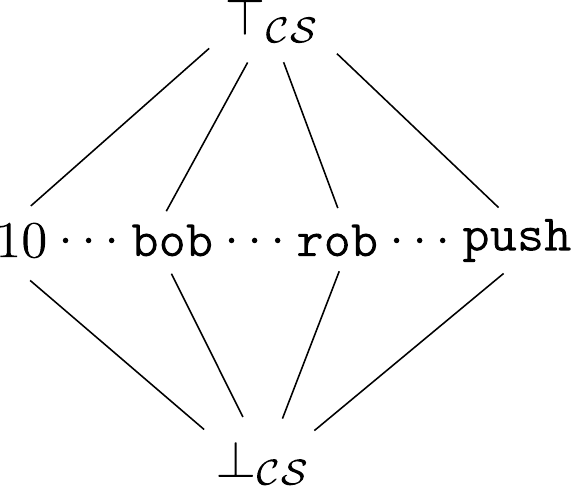}
        \caption{Lattice for \csdom \label{fig:csLattice}}
    \end{subfigure}
    \caption{Lattices for the \nos and \csdom  domains}
\end{figure}

The JSAI domain is a product of the \csdom and the \nos string domains.

\paragraph{Constant String Domain (\csdom).} The \csdom domain tracks constant strings, i.e., it maintains a single concrete string; if the string is not constant, the abstract value is $\top_{\csdom}$. 

\paragraph{Number-Special-or-Other Domain (\nos).}
This domain is a refinement of the \no domain: in addition to tracking \textsc{NumStr} and \textsc{OtherStr}, it also keeps track of special strings from JavaScript in \textsc{SpecialStr}. \textsc{SpecialStr} allows better analysis of JavaScript programs by  special-casing
JavaScript keywords, e.g., \texttt{length, concat, join, pop, push, shift, sort, splice, reverse, valueOf, toString, indexOf, lastIndexOf, constructor, isPrototypeOf, toLocaleString,	hasOwnProperty,} and \texttt{propertyIsEnumerable}. Concatenating a special string with either another special string or a numeric string always produces an  \textsc{OtherStr} string, which is neither special nor numeric. Additionally, concatenating a special string with an \textsc{OtherStr} string always results in a \textsc{NotNum} string.


\Cref{fig:nosLattice} and \Cref{fig:csLattice} show the lattices for the \csdom, and \nos domains.

\subsubsection{DSL used.}
The DSL $\Lang_\jsai$ is essentially the same as the DSL used by Kalita et al.~ \cite{amurth,amurthArtifact}.

\begin{figure}
\hspace{5mm}
\begin{subfigure}[b]{.5\linewidth}
\begin{minted}[autogobble, fontsize=\small, tabsize=2, escapeinside=||, breaklines, linenos]{java}
|concat$_{\jsai}^\sharp$|(|$\mathtt{arg_1, arg_2}$|) {
	if(|$\mathtt{arg_1.cs} \not\in \{\top_{\csdom},\bot_{\csdom}\}  \land$ $ \mathtt{arg_2.cs} \not\in \{\top_{\csdom},\bot_{\csdom}\} $|)
	{
		|$\mathtt{sset \gets \emptyset}$|
		for(|$\mathtt{x \gets arg_1.cs}$|)
			for(|$\mathtt{y \gets arg_2.cs}$|)
				|$\mathtt{sset \gets sset \cup \{\concat(x, y)\}}$|
		|$\mathtt{out.nos \gets \alpha_{\nos}(sset)}$|
		|$\mathtt{out.cs} \gets \mathtt{\alpha_{\csdom}(sset)}$|
		return out
	} else {
		return |\ \ $\langle\mathtt{\concat_{\csdom}^\sharp(arg_1.cs, arg_2.cs)},$ $\mathtt{\concat_{\nos}^\sharp(arg_1.nos, arg_2.nos)}\rangle$\labelline{code:concatDirectJsai}|
  }
}
\end{minted}
\caption{
  $\Lang_\jsai$-transformer for \concat synthesized by \tool
  \label{fig:concatJSAI}}
\end{subfigure}
\hspace{5mm}
\begin{subfigure}[b]{.48\linewidth}
\centering
\begin{minted}[autogobble, fontsize=\small, tabsize=2, escapeinside=||, breaklines, linenos]{C}
|$\mathtt{toLower_{\jsai}^\sharp}(\mathtt{arg_1})$| {
  if(|$\mathtt{arg_1.cs} \not\in \{\top_{\csdom}, \bot_{\csdom}\}$|) {
    |$\mathtt{sset \gets \emptyset}\labelline{code:toLowJsai1}$|
    for(|$\mathtt{x \gets arg_1.cs}$|)
      |$\mathtt{sset \gets sset \cup \{toLower(x)\}}$|
    |$\mathtt{out.nos \gets \alpha_{\nos}(sset)}$|
    |$\mathtt{out.cs \gets \alpha_{\csdom}(sset)}$|
    return out |$\labelline{code:toLowJsai2}$|
  } else {	
    return |$\langle \mathtt{\abs{toLower}_{\csdom}(arg_1.cs)},$ $\mathtt{\abs{toLower}_{\nos}(arg_1.nos)} \rangle$|
  }
}
\end{minted}
\caption{
  $\Lang_\jsai$-transformer for \tolower synthesized by \tool
       \label{fig:toLowerJSAI}}
\end{subfigure}

\hspace{5mm}
\begin{subfigure}[b]{.48\linewidth}
\vspace{5mm}
    \centering
\begin{minted}[autogobble, fontsize=\small, tabsize=2, escapeinside=||, breaklines, linenos]{C}
|$\mathtt{toUpper_{\jsai}^\sharp}(\mathtt{arg_1})$| {
  if(|$\mathtt{arg_1.cs} \not\in \{\top_{\csdom}, \bot_{\csdom}\}$|) {
    |$\mathtt{sset \gets \emptyset}$|
    for(|$\mathtt{x \gets arg_1.cs}$|)
      |$\mathtt{sset \gets sset \cup \{toUpper(x)\}}$|
    |$\mathtt{out.nos \gets \alpha_{\nos}(sset)}$|
    |$\mathtt{out.cs \gets \alpha_{\csdom}(sset)}$|
    return out
 } else {
   return |$\langle\mathtt{ \abs{toUpper}_{\csdom}(arg_1.cs)},$ $\mathtt{\abs{toUpper}_{\nos}(arg_1.nos)} \rangle$|
 }
}
\end{minted}
    \caption{
  $\Lang_\jsai$-transformer for \toupper synthesized by \tool
        \label{fig:toUpperJSAI}}
\end{subfigure}
\begin{subfigure}[b]{.48\linewidth}
    \centering
\begin{minted}[autogobble, fontsize=\small, tabsize=2, escapeinside=||, breaklines, linenos]{C}
|$\mathtt{\contains_{\mathtt{\jsai}}^\sharp(arg_1, arg_2})$| {
  if(|$\mathtt{arg_1.cs \not\in \{\top_{\csdom},\bot_{\csdom}\}} \land$  $\mathtt{arg_2.cs \not\in \{\top_{\csdom},\bot_{\csdom}\}}$ \labelline{code:containsJsai1}|) {
    for |$x \gets \mathtt{arg_1.cs}$|
      for |$y \gets \mathtt{arg_2.cs}$|
        r |$\gets$| contains(x,y)
    if(r) return BoolTrue
    return BoolFalse |$\labelline{code:containsJsai2}$|
  } else {
    return |$\mathtt{ contains_{\csdom}^\sharp(arg_1.cs, arg_2.cs)}$ $\sqcap$ $\mathtt{contains_{\nos}^\sharp(arg_1.nos, arg_2.nos) }$|
  }
}
\end{minted}
\caption{
  $\Lang_\jsai$-transformer for \contains synthesized by \tool
        \label{fig:containsJSAI}}
\end{subfigure}
\caption{$\Lang_\jsai$-transformers synthesized by \tool (part 1)\label{fig:JSAITransformers1}}
\end{figure}

\begin{figure}
\hspace{5mm}
\begin{subfigure}[b]{.48\linewidth}
\centering
\begin{minted}[autogobble, fontsize=\small, tabsize=2, escapeinside=||, breaklines, linenos]{C}
|$\mathtt{trim_{\mathtt{\jsai}}^\sharp(arg_1})$| {
  if(|$\mathtt{arg_1.cs \not\in \{\top_{\csdom},\bot_{\csdom}\}}$|) {
    |$\mathtt{sset \gets \emptyset}\labelline{code:trimJsai1}$|
    for(|$\mathtt{x \gets arg_1.cs}$|)
      |$\mathtt{sset \gets sset \cup \{trim(x)\}}\labelline{code:trimJsaiAp}$|
    |$\mathtt{out.nos \gets \alpha_{\nos}(sset)}$|
    |$\mathtt{out.cs \gets \alpha_{\csdom}(sset)}$|
    return out|$\labelline{code:trimJsai2}$|
 } else {
  return |$\langle\mathtt{ \abs{trim}_{\csdom}(arg_1.cs),}$ $\mathtt{ \abs{trim}_{\nos}(arg_1.nos)} \rangle$|
 }
}
\end{minted}
\caption{
  $\Lang_\jsai$-transformer for \trim synthesized by \tool
            \label{fig:trimJSAI}}
\end{subfigure}
\begin{subfigure}[b]{.48\linewidth}
\centering
\begin{minted}[autogobble, fontsize=\small, tabsize=2, escapeinside=||, breaklines, linenos]{C}
|$\mathtt{charAt_{\mathtt{\jsai}}^\sharp(arg_1: \jsai, pos: NUM})$| {
  if(|$\mathtt{arg_1.cs} \not\in \{\top_{\csdom}, \bot_{\csdom}\} $| 
        |$\land\ \mathtt{pos} \not\in \{\top_{num}, \bot_{num}\}$|) {
    |$\mathtt{sset \gets \emptyset}\labelline{code:charatSafe1}$|
    for(|$\mathtt{x \gets arg_1.cs}$|)
      |$\mathtt{sset \gets sset}\ \cup$ $\mathtt{x.len() \geq pos}$ ? $\mathtt{\{charAt(x, pos)\}: EMPTY} \labelline{code:charAtJsaiAp}$|
    |$\mathtt{out.nos \gets \alpha_{\nos}(sset)}$|
    |$\mathtt{out.cs \gets \alpha_{\csdom}(sset)}$|
    return out |$\labelline{code:charatSafe2}$|
  } else {
    return |$\mathtt{\langle \charat_{\csdom}^\sharp(arg_1.cs, pos)},$ $ \mathtt{\charat_{\nos}^\sharp(arg_1.nos, pos)}\rangle$|
  }
}
\end{minted}
\caption{
  $\Lang_\jsai$-transformer for \charat synthesized by \tool
            \label{fig:charatJSAI}}
\end{subfigure}
\caption{$\Lang_\jsai$-transformers synthesized by \tool (part 2)\label{fig:JSAITransformers}}
\end{figure}

\subsubsection{The \concat operation.}
We show the reduced $\Lang_\jsai$-transformer for \concat synthesized by \tool in \Cref{fig:concatJSAI}.
The manually written $\Lang_\jsai$-transformer for \concat provided by \safestr is semantically equivalent to that synthesized by \tool.

\subsubsection{The \tolower, \toupper,  \contains, \trim, \charat operations.}
We show the reduced $\Lang_\jsai$-transformers for these operations that were synthesized by \tool in \Cref{fig:toLowerJSAI}, \Cref{fig:toUpperJSAI},\Cref{fig:containsJSAI},  \Cref{fig:trimJSAI},  and \Cref{fig:charatJSAI}, respectively. 
In all these cases, the implementation available in \safestr is essentially the direct product. The reduced transformers synthesized by \tool for these operations (except \contains) are more precise. The reasons for improved precision are similar to those already discussed for the \safe domain; for brevity, we omit a detailed discussion of these transformers.

\begin{table}[t]
    \centering
    \caption{Timings to synthesize transformers for string domains in seconds\label{tab:string_timings}}
    \begin{tabular}{|c|c|c|c|c|c|c|}
        \hline
       \diagbox{\bf Dom.}{\bf Func.} & \textbf{concat} & \textbf{contains} & \textbf{toLower} & \textbf{toUpper} & \textbf{trim} & \textbf{charAt}\\
        \hline
        \hline
        \textbf{\safe} & 127 & 218 & 86 & 70 & 134 & 126\\
        \hline
        \textbf{\jsai} &  57 & 21 & 19 & 9 & 11 & 13 \\
        \hline
    \end{tabular}
\end{table}

\paragraph{\bf Concluding Remarks for \safe and \jsai.}
We provide the time taken by \tool to synthesize the reduced transformers in the \safe and \jsai domains in \Cref{tab:string_timings}, which shows that \tool can synthesize reduced transformers for real-world verification engines in a reasonable time. We are planning to lodge a pull request on the \safestr repository to provide the improved transformers automatically synthesized by \tool. 
\section{Related Work}


Program-synthesis techniques are widely accepted in the community and have been used in many different areas of computer science.
CEGIS~\cite{sketch} is one of the popular program-synthesis strategies.
The concept of using a dual CEGIS loop to generate \textit{positive} and \textit{negative} examples for synthesis shows success in synthesizing abstract transformers for concrete operation~\cite{amurth}, as well as the synthesis of specifications~\cite{OOPSLA23:PDR}. 

The research that motivated our work focuses on synthesizing most-precise abstract transformers using a user-specified DSL~\cite{amurth}. The core algorithm of synthesizing abstract transformers is driven by dual CEGIS loops, generating positive and negative examples.
Although \amurth proved capable of synthesizing abstract transformers, it failed to synthesize reduced transformers, even with a significantly large timeout threshold (10 hours).
The reason behind the failure of the synthesis procedure is that \amurth
needs to synthesize the transformers for all of the domains simultaneously, which blows up the search space in which a best transformer is to be found.

Prior to \amurth there have been many works~\cite{VMCAI:RSY04,DBLP:conf/vmcai/KingS10,DBLP:conf/sas/ThakurER12,DBLP:conf/cav/ThakurR12,ThakurLLR15,VMCAI:Reps16} that create best abstract transformers for various abstract-interpretation frameworks with a variety of different requirements.
Reps and Thakur \cite[\S5.2]{VMCAI:Reps16} describe how such techniques can be used to perform semantic reduction in a product domain.
Work by X. Wang et al.~\cite{CAV:Wang18} describe a method for learning abstract transformers for a given abstract domain within a specific language of fixed predicates over affine expressions.
Recent work by J.\ Wang et al.~\cite{ICSE:Wang21} describes another program synthesis-based technique, which uses learned predicates to synthesize a sound abstract transformer; unlike \amurth and \tool, it only focuses on soundness and does not have a mechanism to check the precision of the synthesized transformers. 
\section{Conclusion}
\label{sec:conclude}

Even with over four decades of use of abstract-interpretation-based verification tools, designing sound and precise abstract transformers has remained a challenge.
Transformers for reduced-product domains are even more challenging, because each component transformer now has access to abstract input values from other component domains, and the component transformers must \textit{cooperate} to produce a sound and maximally precise reduced transformer.
Because directly synthesizing all the component transformers for the product domain is not practical, the algorithm presented in this paper iteratively synthesizes the component transformers, one-by-one---each synthesis of a component transformer $\absr{f}_i$ being \textit{conditioned} on all other component transformers---until a sound and maximally 1-precise $\Lang$-transformer is obtained. We used \tool, an implementation of our algorithm, to synthesize reduced-product abstract transformers for two string product domains, SAFE and JSAI, available within the \safestr JavaScript-analysis framework. \tool synthesizes more precise transformers for four of the six supported string operations, for both the SAFE and JSAI domains.

This work is in the same direction as \amurth~\cite{amurth}, which proposed an algorithm for synthesizing abstract transformers for single abstract domains.
We believe that this direction of work---aimed at reducing the effort required to implement key components of verification engines---would not only make verification tools more easily available for new languages,
including small domain-specific languages, but also improve
user-confidence in the judgements reached by verification tools.
In the future, we are interested in applying \tool with sophisticated reduced-product domains for analysis of popular intermediate representations like LLVM bytecode.
The large number of opcodes available, and the sometimes complex semantics of LLVM instructions, seems to make LLVM
a perfect use-case for \tool.


\section*{Acknowledgments}
We thank the anonymous reviewers for their valuable input.
We are thankful to Intel for supporting the first author via the Intel India Research Fellowship Program for doctoral students.

\bibliographystyle{splncs04}
\bibliography{refs}

\end{document}